\documentstyle[aps,prl,epsfig,multicol,amssymb]{revtex}
\epsfclipon

\begin{document}

\newcommand \be {\begin{equation}}
\newcommand \ee {\end{equation}}
\newcommand \bea {\begin{eqnarray}}
\newcommand \eea {\end{eqnarray}}
\newcommand \nn {\nonumber}
\newcommand \la {\langle}
\newcommand \ra {\rangle}
\newcommand \rat {\rangle_{\tau}}
\newcommand \raw {\rangle_{_W}}

\title{\bf Sub-diffusion and localization in the one dimensional trap model}
\author{E.M. Bertin, J.-P. Bouchaud}
\address{\it Commissariat \`a l'\'Energie Atomique,
Service de Physique de l'\'Etat Condens\'e\\
91191 Gif-sur-Yvette Cedex, France}

\maketitle

\begin{abstract}
We study a one dimensional generalization of the exponential trap model 
using both numerical simulations and analytical approximations.
We obtain the asymptotic shape of the average diffusion front in the sub-diffusive phase. 
Our central result concerns the localization properties. We find the dynamical participation 
ratios to be finite, but different from their equilibrium counterparts. Therefore, 
the idea of a partial equilibrium within the limited region of space explored by the walk
is not exact, even for long times where each site is visited a very large number of times. 
We discuss the physical origin of this discrepancy, and characterize the full 
distribution of dynamical weights. We also study two different two-time correlation 
functions, which exhibit different aging properties: one is `sub-aging' whereas the 
other one shows `full aging'; therefore two diverging time scales appear in this model. 
We give intuitive arguments and simple analytical approximations that account 
for these differences, and obtain new predictions for the asymptotic (short time and
long time) behaviour of the scaling functions. Finally, we discuss the issue of multiple 
time scalings in this model.\\
\\
PACS numbers: 75.10.Nr, 05.20.-y, 02.50.-r
\end{abstract}

\begin{multicols}{2}

\section{Introduction}

A lot of efforts have been devoted to the theoretical study 
of aging phenomena in the past decades \cite{Struick,Vincent,Review}. 
Spin glass models, which exhibit a very rich phenomenology, have been widely 
studied theoretically both using analytical techniques for the mean field models, 
or by numerical simulations in the finite dimensional cases. 
Besides these microscopic spin models, a simpler but phenomenological picture, 
the `trap model', has been proposed in order to describe the phase space dynamics 
in a coarse-grained manner \cite{Bou92}. This model seems to capture, at least 
qualitatively, some of the physics involved in the aging dynamics of several 
systems beyond spin glasses, such as fragile glasses \cite{Heuer,Reichman,Berthier},
soft glassy materials \cite{Cates,Sollich}, granular materials \cite{Head}, pinning 
of extended defects (such as domain walls, vortices, etc.) \cite{BBM}. 
This trap model has been studied mainly in its fully connected (or `mean field')
version \cite{Dyre,Dean,Monthus}, which has recently been shown to describe exactly 
the long time dynamics of the Random Energy Model when the distribution of trap depth
is exponential \cite{BenArous}. This version of the mean field model already exhibits a number of 
interesting features, such as a transition between a stationary, `liquid' phase, and 
an aging `glassy' phase, violation of 
the Fluctuation Dissipation Relation \cite{FieldingSollich}, and dynamical ultrametricity
\cite{CuKu,Bertin}. In the glassy phase, the dynamics is strongly intermittent, since most
of the time nothing happens, whereas the active periods appear in bursts which become less
and less frequent as time elapses. Several recent experiments suggest that such an intermittency is indeed
present in glassy systems \cite{Israeloff,Cipelletti,Ciliberto}, or in atomic physics 
\cite{Bardou,Barkai,Dahan}.

The finite dimensional generalization of this model has already been studied many years
ago \cite{Alex81,Alexander,Machta,PhysRep}, but only one time quantities (not well suited to study
aging) were considered. These aging properties were addressed only recently in 
\cite{Monthus,Maass}, and, from a more rigorous point of view, in \cite{Isopi,BenArous2}.
One expects that in dimensions $d >2$, the trap model will have properties qualitatively 
similar to the fully connected case, since each site is visited by the walk a finite 
number of times. In lower dimensions $d \leq 2$, the correlations induced by the multiple
visits of the walks to a given site is expected to lead to qualitative changes. It was
for example shown in \cite{Maass} that some quantities exhibit {\it sub-aging} properties, i.e.
decay on a time scale that scale with the waiting time $t_w$ as $t_w^\nu$ with $\nu < 1$.
Because of the limited number of accessible sites, one might also expect interesting
properties such as {\it dynamical localization}, which means that there is a finite probability
that $k$ independent particles sit on the very same site, even after a very long waiting time
$t_w$. Such a dynamical localization was first established by Golosov in the 
context of the Sinai model \cite{Golosov} and extended to the biased case in \cite{Compte}, 
and more recently proven rigorously for the one dimensional trap model in \cite{Isopi}. 

In this paper, we present a detailed study of the one dimensional (non biased) trap model, 
using both numerical simulations and analytical approximations. In the first 
section, we focus on the scaling form of the average ``diffusion front'' 
$\langle p(x,t) \rangle$ in the sub-diffusive, non Gaussian phase, for which no
analytical results are (to our knowledge) available. We present 
some scaling arguments and approximation schemes to account for our numerical data. 
We then discuss the idea of partial equilibrium in this model, which can be explored
in details through the distribution of dynamical weights. The moments of this distribution 
are the usual `participation ratios' that characterize the localization properties of
the measure. Perhaps surprisingly, these localization indicators are indeed finite (as 
first shown in \cite{Isopi}), but different from their static counterparts. We discuss in detail
the origin of this difference, and try to characterize quantitatively the distribution of 
dynamical weights. In the last section, we study the aging behaviour of two different
correlation functions, which exhibit different scaling properties, meaning that two
different time scales, $t_w^\nu$ and $t_w$, appear in this model. We again develop 
intuitive arguments and simple analytical approximations to understand these differences, 
and obtain new predictions for the asymptotic behaviour of the scaling functions, which 
are found to be in excellent agreement with the numerics. Finally, we discuss the 
possible existence of multiple time scalings in this model (as can happen in spin glasses
\cite{CuKu} or in a generalization of the trap model \cite{Maass}). 

\section{The one dimensional trap model. Anomalous diffusion}

\subsection{Definition of the model}

Consider a one dimensional lattice, and define on each site $i$ a quenched random variable 
$E_i>0$ chosen from a distribution $\rho(E)$. $E_i$ has to be interpreted as the energy 
barrier that the particle (the walker) has to overcome in order leave the site. The 
dynamics is chosen to be activated with temperature $T$, which means that the escape 
rate $w_i$ of site $i$ is given by $w_i=\Gamma_0 e^{-E_i/T}$, where $\Gamma_0$ is a 
microscopic frequency scale. Once that particle has escaped the trap, 
it chooses one of the two neighbouring sites, with probability $q_-$ for the left one and 
$q_+=1-q_-$ for the right one. The `directed' case $q_+=1$ is quite simple to 
analyse analytically, since each since is visited once -- see \cite{LeDou,PhysRep,Compte}.
The case $q_{+}=1/2 $ that we study in the following is much more subtle since each
site is visited a large number of times, inducing long range correlations in the 
hopping rates seen by the walker. (Note that as soon as $q_{+} \neq 1/2$, one
expects the large time properties of the walk to be the same as in the fully directed case 
\cite{Compte}.)

For the purpose of heuristic arguments and Monte-Carlo studies, it is interesting to
study the trapping time $\tau$ of the particle on each site. 
Once the transition rates $w_i$ are given, $\tau$ is a random variable with a (site dependent) distribution $p_i(\tau)=w_i\, e^{-w_i \tau}$, of mean $\tau_i=w_i^{-1}$. 
If we choose an exponential density of trap depths, $\rho(E) = \frac{1}{T_g}e^{-E/T_g}$, 
then the distribution of $\tau_i$'s over the different sites is a power law:
\be \label{time_dist}
\psi(\tau) = \frac{\mu \hat{\tau}_0^{\mu}}{\tau^{1+\mu}}  \qquad (\tau \geq \hat{\tau}_0),
\ee
where $\mu=T/T_g$ is the reduced temperature, and $\hat{\tau}_0 \equiv \Gamma_0^{-1}$. For $T>T_g$, this distribution has a finite average value $\langle \tau \rangle = \hat{\tau}_0/(\mu-1)$. This corresponds to 
usual diffusion and stationary dynamics, with a diffusion constant $D=a^2/\langle \tau \rangle$,
where $a$ is the lattice spacing. On the contrary, for $T \leq T_g$, the first moment of the distribution diverges, diffusion becomes anomalous and aging effects are 
expected. A dynamical phase transition takes place at $T_g$, as in the fully connected model. However, new properties emerging from the non trivial spatial structure of the model 
are expected.

\subsection{Disorder induced sub-diffusion: A scaling argument}

We first give a simple scaling argument (proposed in \cite{Alex81,Bou87,PhysRep}) that yields a sub-diffusive behaviour for the one dimensional trap model introduced above. In the following, we shall take $a$ is the unit of length, as well as $\Gamma_0^{-1}$ as the time unit. Roughly speaking, a typical random walk starting from a given initial site has visited, after $N$ steps, of the order of $\sqrt{N}$ sites (which implies a typical displacement $\xi \sim \sqrt{N}$). Each site is 
visited around $\sqrt{N}$ times. So the time $t$ elapsed can be written as:
\be
t \sim \sqrt{N}\, \sum_{i=-\sqrt{N}}^{\sqrt{N}} \tau_i.
\ee
Since the sum of $M$ independent random variables distributed according to Eq.~(\ref{time_dist}) grows as $M^{\frac{1}{\mu}}$, we get:
\be
t \sim \sqrt{N}^{1+\frac{1}{\mu}} \sim \xi^{1+\frac{1}{\mu}}.
\ee
Inverting this relation leads to the following sub-diffusive behaviour:
\be \label{xi-t}
\xi(t) \sim t^{\frac{\mu}{1+\mu}}.
\ee
This result was also obtained by Machta \cite{Machta}, using Real Space Renormalization
Group arguments. The same behaviour also holds for the {\it random barrier} model
with a broad distribution of barrier heights which, in one dimension, is expected to be equivalent to the trap model, as far as diffusion properties are concerned \cite{PhysRep}. For this model, the average probability of being
on the initial site can be exactly computed, and decays as $1/t^{\frac{\mu}{1+\mu}} = 
1/\xi(t)$, in agreement with the above result. The exponent $\frac{\mu}{1+\mu}$ is also 
in very good agreement with numerical results \cite{Maass}. 

The case $\mu=1$ is special since logarithmic corrections come into play. Extending the
above argument leads to:
\be \label{xi-t-mu=1}
\xi(t) \sim \sqrt{\frac{t}{\ln t}} \qquad (\mu=1),
\ee 
whereas for $\mu > 1$ one recovers $\xi(t) \sim \sqrt{t}$.

Calling $p(x=i\,a,t)=P_i(t)$ the probability to be at a distance $x$ from the starting point
after time $t$, one expects the disordered average diffusion front 
$\langle p(x,t) \rangle_\tau$ to 
take for large times the following scaling form:
\be \label{scal_pxt}
\la p(x,t) \rat = \frac{1}{\xi(t)} \, f\left(\frac{x}{\xi(t)}\right),
\ee
where $\xi(t)$ is given by Eq.~(\ref{xi-t}), and $\la \ldots \rat$ stands for the average over the quenched trapping times $\tau_i$. However, the full scaling function $f(.)$ is,
to our knowledge, not known. Only the value of $f(0)$ in the dual `barrier' model 
was obtained in \cite{Alexander}. Before studying more subtle issues, we have 
investigated this question both analytically and numerically.  
 
\subsection{The average diffusion front}

For a system without disorder, or in the case $\mu > 1$ where the average trapping time $\langle \tau \rangle$ is finite, the Central Limit Theorem tells us immediately 
that the diffusion front becomes Gaussian at large times: 
\be
\langle p(x,t) \rangle_\tau = \frac{1}{\sqrt{2\pi Dt}}\, \exp \left(-\frac{x^2}{2Dt}\right)
\ee
where $D=a^2/\langle \tau \rangle$ is the diffusion constant, $a$ being the lattice spacing, so that $D \propto (\mu -1)$ when $\mu \to 1^+$. In the case $\mu \leq 1$, a modified space-time scaling is expected, as argued in the previous subsection, as well as a non Gaussian diffusion front.

\begin{figure}
\centerline{
\epsfxsize = 8.5cm
\epsfbox{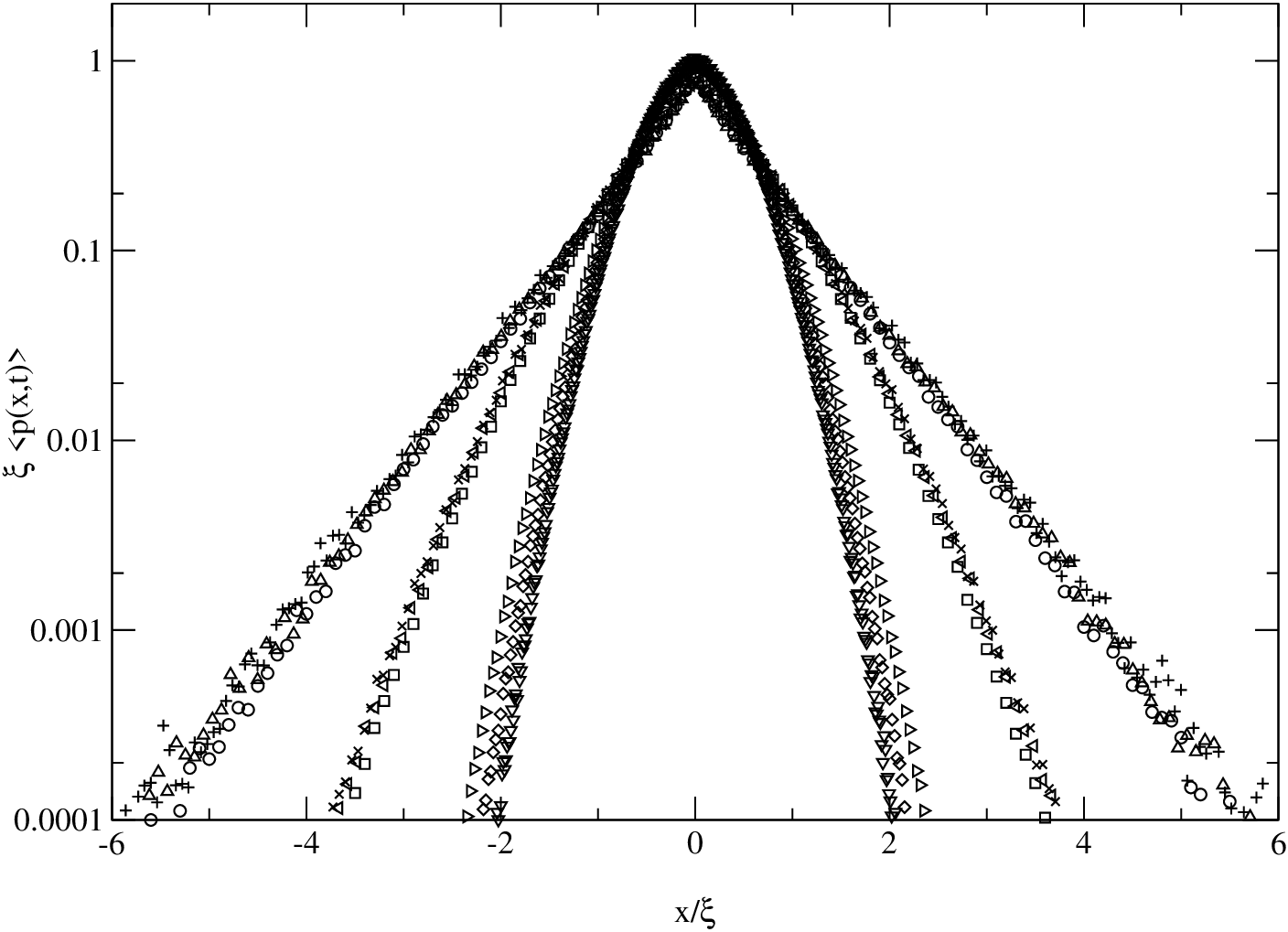}
}
\vskip 0.5 cm
\caption{\sl Plot of $\xi(t)\, \la p(x,t) \ra$ versus $x/\xi(t)$ for different temperatures and different times. Data were obtained by Monte-Carlo simulations on the model with fixed trapping times. Upper curves: $\mu=0.2$, and $t=10^6\, (\circ)$, $10^8\, (\vartriangle)$, $10^{10}\, (+)$; middle curves: $\mu=0.5$ and $t=10^3\, (\Box)$, $10^4\, (\vartriangleleft)$, $10^5\, (\times)$; lower curves: $\mu=0.9$ and $t=10^3\, (\vartriangleright)$, $10^4\, (\Diamond)$, $10^5\, (\triangledown)$.} 
\label{pxt-fig}
\end{figure}

We have developed simple approximation schemes (that we expect to become exact in the 
limits $\mu \to 1$ and $t \to \infty$) to compute $\la p(x,t) \rat$
analytically. The calculations are reported in Appendix A. We find that 
$\la p(x,t) \rat$ can indeed be written as Eq.~(\ref{scal_pxt}) with 
$\xi(t)$ given by Eq.~(\ref{xi-t}). The asymptotic shape of $f(\zeta={x}/{\xi(t)})$ 
can furthermore be computed in the limits $\zeta \to 0$ and $\zeta \to \infty$.
We find:
\bea \label{px1}
f(\zeta) &\approx& f_\infty |\zeta|^{\alpha} \exp(-b |\zeta|^{\beta}) \qquad |\zeta| \to \infty\\
f(\zeta) &\approx& f_0 - f_1 |\zeta|^{\mu} - f_2 |\zeta|^{\gamma} \qquad |\zeta| \to 0,
\eea
where $\alpha=(\mu-1)/2$, $\beta=1+\mu$ and $\gamma=\min(2,1+2\mu)$.
The constants $f_0,f_1,f_2,f_\infty$ and $b$ are $\mu$ dependent numbers that we can also compute 
in the Appendix (see Eqs~(\ref{factor1}) and (\ref{factor2})).  

In the case $\mu=1$, we find, using the same approximation (which we now believe is exact), that $\zeta$ is given by 
$x \sqrt{\ln x/t} \approx x \sqrt{\frac{1}{2}\ln t/t}$, and that $f(\zeta)$ is exactly Gaussian, as for the `normal' case $\mu > 1$. More precisely, one finds for the diffusion front, in the limit $t \to \infty$:
\be \label{px-crit}
\la p(x,t) \rat \simeq \sqrt{\frac{\ln t}{4\pi t}} \exp \left(-\frac{x^2}{4t}\ln t \right)
\ee
We have tested numerically the validity of the scaling relation (\ref{scal_pxt}), for several values of $\mu$. The plot of $\xi(t)\, \la p(x,t) \ra$ as a function of ${x}/{\xi(t)}$, for different values of $t$ shows a rather good collapse. The curves collapse well for $\mu=0.2$ and $0.5$, even if for $\mu=0.2$ data is more noisy. However, finite time corrections become stronger as $\mu$ approaches $1$.
This is expected: sub-leading corrections to scaling can be shown to become negligible 
only in the limit where $t^{\frac{1-\mu}{1+\mu}} \gg 1$. For $\mu=0.9$ and $t=10^5$, however, this
parameter is only $\approx 1.8$. Therefore, we expect that the $\mu=0.9$ data will
actually be strongly affected by the vicinity of $\mu=1$, which plays the role of a critical point.

\begin{figure}
\centerline{
\epsfxsize = 8.5cm
\epsfbox{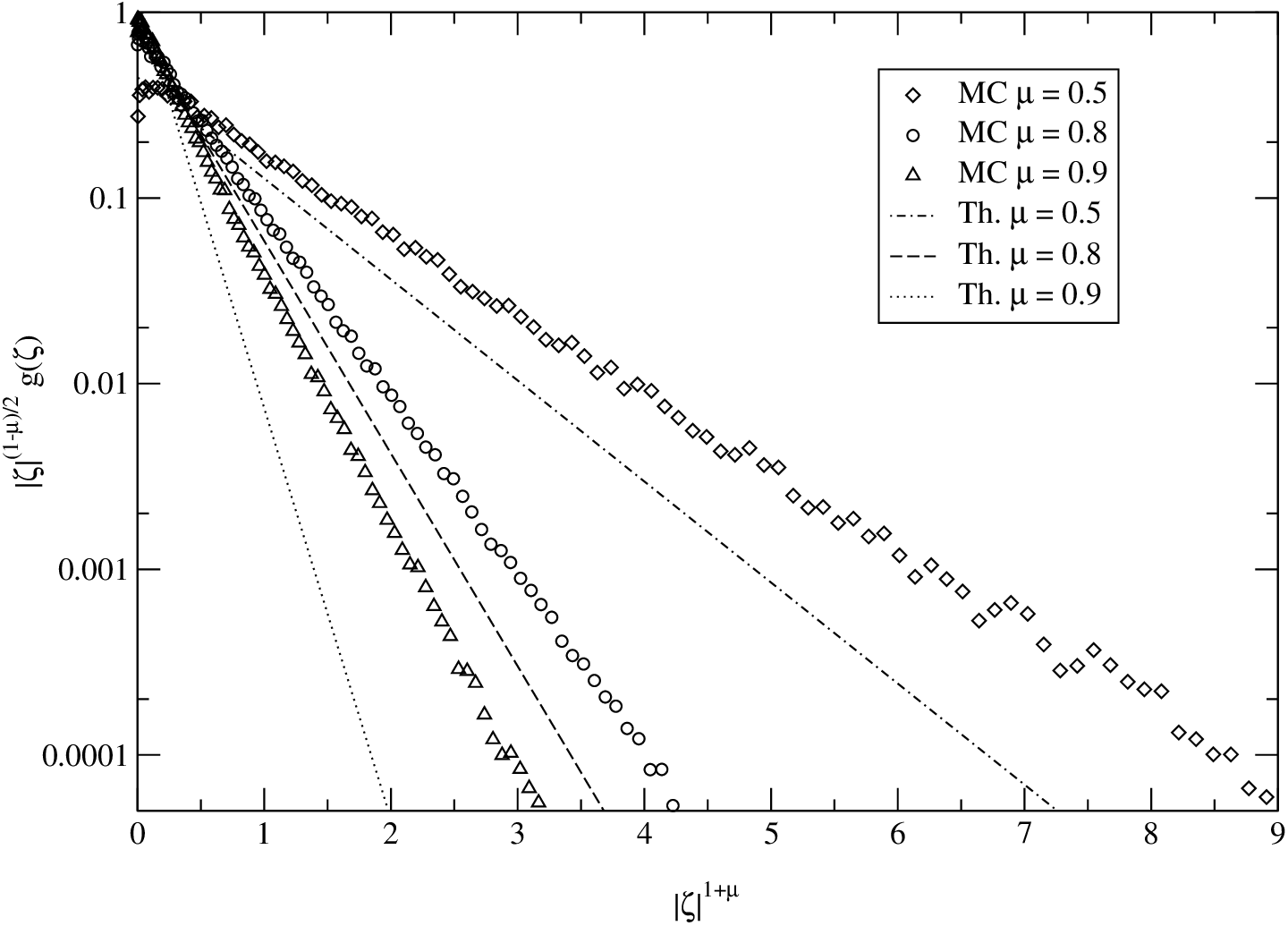}
}
\vskip 0.5 cm
\caption{\sl Plot of $|\zeta|^{(1-\mu)/2} g(\zeta)$ versus $|\zeta|^{1+\mu}$ for $\mu=0.5\, (\Diamond)$, $0.8\, (\circ)$ and $0.9\, (\vartriangle)$, obtained by an infinite time extrapolation of the Monte-Carlo data. The analytical prediction, using the approximation valid for $\mu$ close to $1$, is also shown for the same values of $\mu$ (lines). The predicted exponent $|\zeta|^{1+\mu}$ is in good agreement with the numerics, since data appear to be linear in this representation. The prediction for $b$ -- see Eq.~(\ref{px1}) -- is in best agreement with the Monte-Carlo data for $\mu=0.8$.
}
\label{px-fig}
\end{figure}

In Fig.~\ref{px-fig} we show the scaling functions for $\mu=0.5$, $0.8$ and $0.9$ 
obtained by extrapolating to $t = \infty$ the scaling curves obtained at finite $t$. We actually plot $f(\zeta)/\zeta^\alpha$ as a function of $\zeta^\beta$ 
in a semi-log plot, in order to test directly the asymptotic form given by Eq.~(\ref{px1}).
Note that the approximation is supposed to be valid only for $\mu$ close to $1$, but seems to work well even for rather small values of $\mu$, like $\mu=0.5$. Also shown are the analytic predictions, with the computed numerical values of the constants $f_\infty$ and $b$. We see that the agreement is quite reasonable, and
actually suggests that the value $\beta=1+\mu$ is probably exact. For $\mu=0.9$, critical corrections become important and the predicted slope is not as good as for $\mu=0.8$, but the exponent $1+\mu$ seems to be correct.

\begin{figure}
\centerline{
\epsfxsize = 8.5cm
\epsfbox{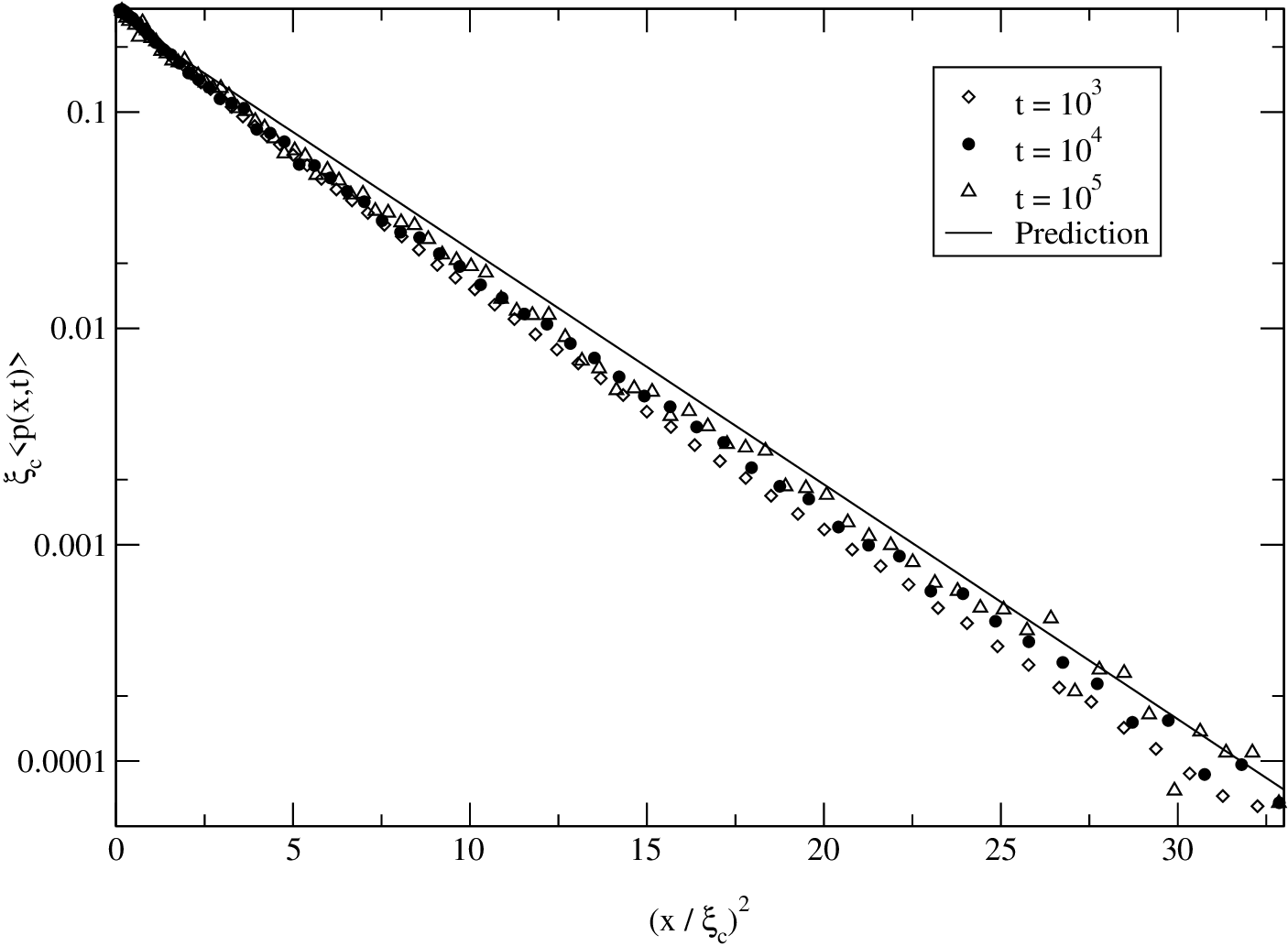}
}
\vskip 0.5 cm
\caption{\sl Plot of $\xi_c(t) \la p(x,t) \rat$ versus $[x/\xi_c(t)]^2$ for $\mu=1$ and times $t=10^3$ ($\diamond$), $10^4$ ($\bullet$) and $10^5$ ($\vartriangle$); $\xi_c(t)$ is the critical coherence length, defined as $\xi_c(t)=[t/\ln(\Gamma t)]^{\frac{1}{2}}$, where $\Gamma$ is fitted on the $t=10^4$ curve. The data points seem to converge at large times towards the asymptotic Gaussian form predicted by Eq.~(\ref{px-crit}) (line).}
\label{px-crit-fig}
\end{figure}

Data corresponding to $\mu=1$ are shown in Fig.~\ref{px-crit-fig}. It is actually necessary to take
into account a finite time correction in this case, replacing $\ln t$ by $\ln (\Gamma t)$, where $\Gamma$ is an unknown constant which has to be fitted.
 This correction is natural, as can be shown by the structure of the sub-leading terms. Fitting $\Gamma$ on data corresponding to $t=10^4$ leads to $\Gamma \simeq 1.64$. One sees that for large $x$, the slope is then found to be the same for the three times simulated.

We have also tested the small $\zeta$ region. For $\mu = 0.5$, the $\sqrt{|\zeta|}$
singularity predicted by our approximation -- see Eq.~(\ref{px1}) -- is rather convincing.
However, as $\mu$ increases towards $1$, the coefficient $f_1$ of $|\zeta|^\mu$ 
decreases towards zero. The next leading term becomes important and one indeed observes an effective singularity with an exponent intermediate between $\mu$ and $2$: we find this
exponent to be $\approx 1.6$ for $\mu=0.8$ and $\approx 1.8$ for $\mu=0.9$.

\section{`Partial Equilibrium' and Localization} 

The one dimensional diffusion problem is interesting because, as mentioned above, 
each site is visited by the walk a large number of times. A natural idea is therefore 
that at time $t$, the probability $P_i(t)$ to find the particle at site $i$ should be 
very similar to the {\it equilibrium} distribution restricted to
an interval of finite length $\propto \xi(t)$. More precisely, we can expect that $P_i(t)$ can be written on the following `quasi-equilibrium' form $P_i^{qe}(t)$:
\bea \label{piwithform}
P_i(t) &\approx& P_i^{qe}(t) = \frac{g_i(t)}{Z} \, e^{E_i/T} \\ \nonumber
Z &=& \sum_{i=-\infty}^{\infty} g_i(t)\, e^{E_i/T},
\eea
where the `form factors' $g_i(t)$ are slowly varying and decay on the scale of $\xi(t)$. 
(Note that the energy barrier $E_i>0$ is the opposite of the energy of the site.)
This idea of `partial
equilibrium' is actually quite general and is often advocated in the context of glassy 
dynamics: although the system is out of equilibrium, one may think of its state at time $t$
as of a partial equilibrium restricted to the region of phase space that it has explored 
up to time $t$, see e.g. \cite{Yoshino,FranzVirasoro,ABarrat}. This idea of partial equilibrium was introduced and used quantitatively in the context of random walk models in \cite{Maass}. 

In this section, we want to discuss this issue in some details. It turns out that the full
statistics of $P_i^{qe}$ can be worked out in the limit where $\xi(t) \to \infty$, 
and can be compared to the corresponding statistics of $P_i(t)$ that we determine 
numerically.  Perhaps surprisingly, we find that these statistics differ significantly
even in the long time limit, meaning that the out of equilibrium problem never
approaches a `quasi-equilibrium' regime.

\subsection{Participation ratios and localization}

In order to investigate the statistical property of a random probability measure (such as $P_i(t)$
or $P_i^{qe}$), one can introduce the following distribution:
\be\label{varphi-def}
\varphi (P) = \la \, \sum_i P\, \delta(P-P_i) \,\rat \qquad 0<P<1
\ee
which is defined in such a way as to give a small weight to the very large number of sites with small energies, in order to evidence the statistics of the deeper traps present in the system. This distribution $\varphi(P)$ is normalized, since:
\be
\int_0^1 \varphi(P) dP = \la\, \sum_i P_i \,\rat = 1
\ee
The moments of this distribution are related to the so-called inverse participation ratios $Y_k$:
\be
\int_0^1 P^{k-1} \varphi(P) dP = \la\, \sum_i P_i^{k} \,\rat = Y_k
\ee
These participation ratios have an interesting interpretation: if $Y_k$ remains finite for 
$k > 1$ as the number of terms in the sum diverges, 
one speaks of {\it localization}, since a finite fraction of the number of particles
remain concentrated on a finite number of sites, even in the limit of an infinite number of available sites. The $Y_k$ where introduced in the context of electronic localization \cite{Mirlin} and
in spin glass theory \cite{MPV}, and studied in several other problems \cite{Derrida,TFR}. Note that in the limit $k \to 1$,
$(Y_k-1)/(1-k)$ becomes the statistical entropy of the measure $P_i$.

In equilibrium, and for integer values of $k$, $Y_k$ can be interpreted as follows: suppose one chooses at random $k$ particles with their corresponding equilibrium weight, $Y_k$ is the probability to find them
all at the same site. Correspondingly, for the out of equilibrium situation, $Y_k$ is the
probability to find $k$ particles (that all started at the same site) clustered together
on the same site at time $t$. Obviously $Y_k$ can only be non zero if some effective 
attraction exists between the particles. In the case of disordered systems, this attraction
is induced by the disordered environment, where (non interacting) particles condense into particularly favorable sites.

For the problem at hand, the quasi-equilibrium value $Y_k^{qe}$ of the participation ratios can be computed, using for instance auxiliary integrals \cite{Derrida} (see also \cite{Mez84,BouMe97}). 
Interestingly, the detailed shape of the `form factors' $g_i$ in Eq.~(\ref{piwithform})
does not matter, and the quasi-equilibrium results $Y_k^{qe}$ coincide with the equilibrium values $Y_k^{eq}$.
For $\mu \geq 1$, $Y_k^{eq}$ tends to $0$ when $\xi(t)$ goes to $\infty$, 
whereas for $\mu<1$ it converges for large $\xi(t)$ to a finite value:
\be \label{Ykeq}
Y_k^{eq}=\frac{\Gamma(k-\mu)}{\Gamma(k)\,\Gamma(1-\mu)},
\ee
identical to that found in the Random Energy Model. This means that in the low temperature phase $T < T_g$, the equilibrium measure localizes over a finite set of sites. The corresponding equilibrium distribution of weights is given by:
\be
\varphi_{eq}(P) = \frac{1}{\Gamma(1-\mu)\,\Gamma(\mu)}\, P^{-\mu} (1-P)^{\mu-1}
\ee

The precise question we want to ask, in order to test the partial equilibrium idea, is
whether or not the dynamical $Y_k(t)$ approach, in the long time limit, the equilibrium $Y_k^{eq}$, computed for a large system size $L$. This can be also seen as a question about the
commutation of the two limits $t \to \infty$ and $L \to \infty$ (see e.g. \cite{Review}). The equilibrium case corresponds to taking $t \to \infty$ first, at fixed $L$, and then taking $L$ to infinity.
The out of equilibrium case, on the other hand, corresponds to taking $L = \infty$ from the outset and let $t \to \infty$.

\subsection{Dynamical localization and weak ergodicity breaking}

Let us now turn to the dynamical localization properties, starting from a localized 
initial condition, $P_i(t=0)=\delta_{i,0}$. It has been shown by Fontes {\it et al} \cite{Isopi} that the random walk process with a diverging local mean trapping time converges, up to a space-time rescaling, to a stationary process. Consequently all spatially integrated (one time) quantities like participation ratios converge to asymptotic values at large time, which are a priori different from the equilibrium ones. Unfortunately, this mathematical approach has not been able yet to predict the corresponding numerical values. 
We have computed numerically, using a simple Monte-Carlo method, the time dependence of 
$Y_k(t)$ for several values of $k$ ($k=2, \frac{5}{2}, 3, \frac{7}{2}, 4$); see Appendix B for technical 
details. Our simulations confirm the convergence of $Y_k(t)$ 
towards a limiting value for large $t$, and show that these asymptotic values are indeed different from the equilibrium ones.

\begin{figure}
\centerline{
\epsfxsize = 8.5cm
\epsfbox{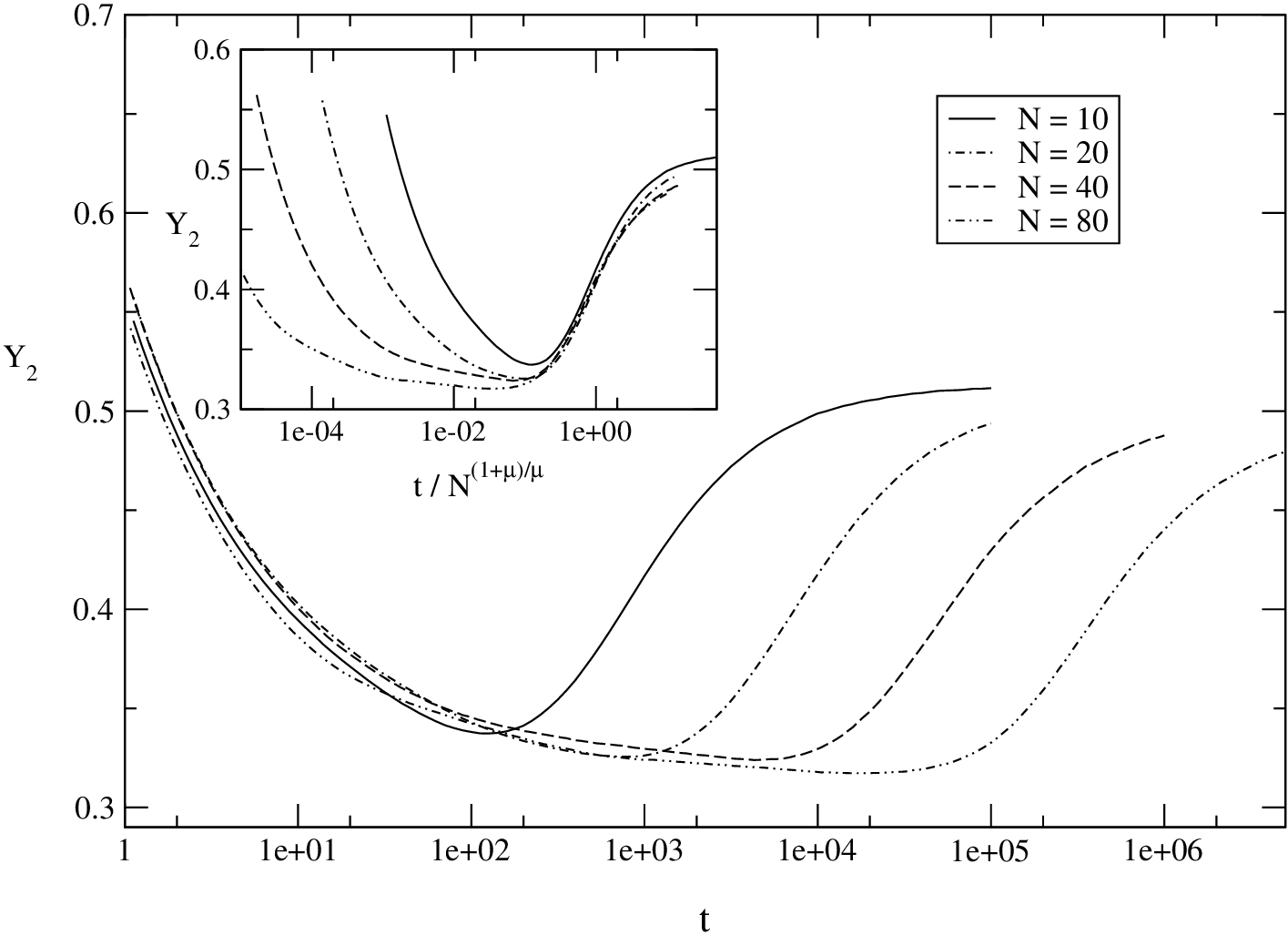}
}
\vskip 0.5 cm
\caption{\sl Test of the convergence of $Y_2(t)$ towards an out of equilibrium value, for $\mu =0.5$: $Y_2(t)$ is plotted for different small sizes $L=2N+1$ of the system, so that equilibration can be reached within simulation time. One can see the onset of a plateau at a value lower that $Y_k^{eq} = 0.5$ (the equilibrium value for $L \to \infty$). The inset shows that the curves collapse if time is rescaled by the equilibration time $t_{erg} \propto N^{(1+\mu)/\mu}$, so that the plateau corresponds to a true out of equilibrium effect, and not to an initial transient.}
\label{conv_Ykt}
\end{figure}

In order to evidence the convergence of $Y_k(t)$ towards different asymptotic values depending on the order of limits
$t \to \infty$ and $L \to \infty$, we have first studied small systems for different sizes $L=2N+1$, in the case $\mu=0.5$. Fig.~\ref{conv_Ykt} shows the onset of a clear plateau at a value $Y_{2}^{dyn}(L)$ smaller than the value predicted by Eq.~(\ref{Ykeq}), $Y_2^{eq} = 0.5$ (for $L \to \infty$), before a cross-over towards the equilibrium regime. Rescaling the time coordinate by a factor $N^{(1+\mu)/\mu}$
(corresponding to the equilibration time $t_{erg}$ of the system), the data collapse rather well, at least in the cross-over region. This shows that the plateau indeed corresponds to the onset of an out of equilibrium steady state regime, when the diffusion length is much smaller than the size of the system. The cross-over appears when the two lengths become comparable.

\begin{figure}
\centerline{
\epsfxsize = 8.5cm
\epsfbox{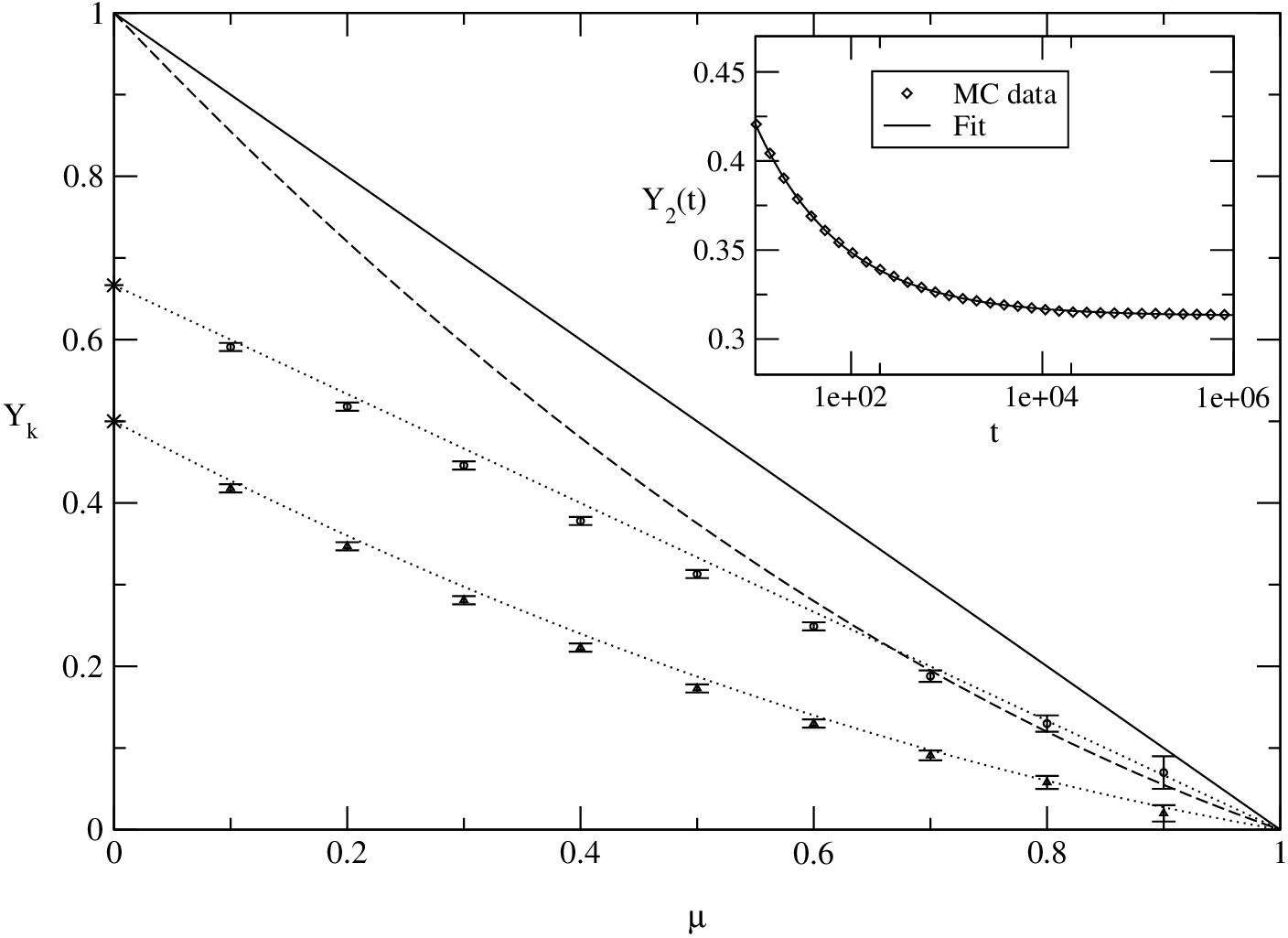}
}
\vskip 0.5 cm
\caption{\sl Plot of $Y_2^{dyn}$ (upper symbols) and $Y_3^{dyn}$ (lower symbols) vs. $\mu$. The equilibrium functions $Y_2^{eq}(\mu)$ (full line) and $Y_3^{eq}(\mu)$ (dashed line) are shown for comparison. One clearly sees that localization is weaker than in the equilibrium situation. In particular, the participation ratios seem to converge to a zero temperature limit which is less than $1$. The stars and the dotted lines correspond to the prediction of a simple model given below, Eq.~(\ref{ykstar}). Inset: fit of $Y_2(t)$ using the functional form $Y_2(t)=Y_2^{dyn}+A t^{-a}$, for $\mu=0.5$. Only $1$ Monte-Carlo point out of $12$ is shown, for clarity.}
\label{Ykmu}
\end{figure}

In order to study the asymptotic value $Y_{2}^{dyn} \equiv Y_{2}^{dyn}(L \to \infty)$, we have simulated systems 
of very large sizes $L$. However, the temporal convergence of $Y_2(t)$ is very 
slow and some infinite time extrapolation procedure is needed. As illustrated in the inset of Fig.~\ref{Ykmu} for $k=2$ and $\mu=0.5$, we have assumed a power-law
convergence of $Y_k$, of the form $Y_k(t)=Y_k^{dyn}+A t^{-a}$ with $3$ fitting parameters $Y_k^{dyn}$, $A$ and $a$, which was
found to work rather well. However, for $\mu$ close to $1$, the fitting parameter $Y_k^{dyn}$ becomes very sensitive to the choice of the time interval used to fit the data, and as a result, error bars become larger. 

Fig.~\ref{Ykmu} shows the extrapolated $Y_2^{dyn}$ and $Y_3^{dyn}$ 
as a function of $\mu$, and compares it to the equilibrium relation $Y_2^{eq}(\mu)=1-\mu$, 
and $Y_3^{eq}(\mu)=(1-\mu)(1-\mu/2)$. It appears that the dynamical localization is weaker than in equilibrium. In particular, $Y_2^{dyn}$ and $Y_3^{dyn}$ converge to a value smaller than $1$ when $\mu$ goes to $0$. We shall argue below that $Y_k^{dyn}(\mu=0)=2/(k+1)$, whereas 
$Y_k^{eq}(\mu=0) = 1$. In the other limit, $\mu \to 1$, it will also be argued in the next section that $Y_k^{dyn}$ vanishes linearly with $\mu$. This is indeed compatible with the numerical data, although other functional dependence might also be compatible since the error bars are large in this range of $\mu$. We have also shown in Fig.~\ref{Ykmu} the prediction of a simple argument
given in section~\ref{Sect-gen-arg} below, which suggests $Y_k^{dyn}(\mu)=2Y_k^{eq}(\mu)/(k+1)$, which is
in rather good agreement with the numerical results.

Therefore, all the dynamical participation ratios $Y_k^{dyn}$ are different from their 
static counterpart. The relative weights of the different visited sites are
not given by the ratio of their Boltzmann weights. This result is important, since it was shown that, in the Sinai model, equilibrium and dynamical participation ratios indeed coincide \cite{MonthusLeDou}.  An interesting possibility, discussed in the context of glassy systems, would be that the $Y_k^{dyn}$ correspond to an equilibrium measure but at a different effective temperature. We shall discuss below that this is not the case either.

Note that for walks in higher dimensions, $d > 2$, one can show rigorously that 
$Y_k^{dyn}=0$ \cite{BenArous2}, whereas $Y_k^{eq}$ are still given by Eq.~(\ref{Ykeq}).\footnote{The case $d=2$ is marginal, but one still finds that $Y_k^{dyn}=0$
in that case \cite{BenArous2}.} However, in this case, each site is visited a finite number of times, and therefore it could have been expected that the idea of partial equilibrium over the set of visited sites would be quantitatively incorrect (although it is able to reproduce, at least qualitatively, some non trivial dynamical correlation functions -- \cite{Maass} and see below). 
Another solvable case is the one dimensional directed walk, where each site is visited 
once. In this case, $Y_2^{dyn}$ can be computed exactly \cite{Compte} and is found to
be close to, but different from, the equilibrium value $1-\mu$. 

The surprising aspect of our result in one dimension is that each site is visited,
asymptotically, an infinite number of times -- a feature that, at least naively, should 
lead to partial equilibration. 

A different, but related, issue concerns the fraction $f_i(t)$ of the total time $t$ a {\it given} particle has spent in the $i^{th}$ trap, and study the participation ratios of this quantity. In this case, we
have found numerically that the different $Y_k$ {\it are} given by the equilibrium formula, Eq.~(\ref{Ykeq}). Therefore,
we are in a situation where ergodicity is (weakly) broken: the relative time a given particle spends on different
sites has not the same statistics as the relative fraction of particles that are found on the different 
sites at a given instant of time. Such a difference between individual and ensemble measurements have
been emphasized in a different context in \cite{Bardou}, and recently observed experimentally \cite{Dahan}.

\subsection{Analytical calculation of the participation ratio}

Using the same procedure as for $\la p(x,t) \rat$, one could try to
compute $Y_k(t)$ which is given by:
\be
Y_k(t) = \int_{-\infty}^{\infty} dx \la p(x,t)^k \rat
\ee
However, this calculation reveals to be much harder than for the case of $\la p(x,t) \rat$. In Appendix C, we report a simplified calculation in the case $k=2$, which aim is to argue that
$Y_2^{dyn}$ is different from $0$ in the low temperature phase $\mu<1$, whereas it vanishes for $\mu>1$. Although this last result has been rigorously proven in \cite{Isopi}, we want to introduce here a general method that could in principle yield bounds and approximations for $Y_k$ for any $\mu$, and not only for $\mu<1$. We obtain the behaviour of $Y_2^{dyn}$ for $\mu \to 1^-$ and find how $Y_2(t)$ 
vanishes as a function of $t$ for $\mu > 1$. To do this, we introduce a function $R(t,t')$ through:
\be
R(t,t') = \int_{-\infty}^{\infty} dx\, \la p(x,t)\,p(x,t') \rat
\ee
(so that $Y_2(t)=R(t,t)$), as well as its Laplace transform $\hat{R}(s,s')$:
\be
\hat{R}(s,s') = \int_0^{\infty} dt \int_0^{\infty} dt' e^{-st-s't'} R(t,t')
\ee
Using rather crude approximations, we obtain that, in the particular case where $s=s'$ and $\mu < 1$:
\be \label{Zss}
\hat{R}(s,s) \simeq \frac{R_0}{s^2} \qquad s \to 0
\ee
with a finite coefficient $R_0$. 
In order to interpret this result, we assume that $R(t,t')$ obeys, for large $t,t'$, 
a scaling relation of the form:
\be
R(t,t') = Y_2^{dyn} {\cal R}\left(\frac{t}{t'}\right)
\ee
which we have confirmed using numerical simulations. Then one gets:
\be
\hat{R}(s,s) \sim \frac{2 Y_2^{dyn}}{s^2} \int_1^{\infty} du \frac{{\cal R}(u)}{(1+u)^2},
\ee
or, using Eq.~(\ref{Zss}),
\be
Y_2^{dyn} = \frac{R_0}{2\int_1^{\infty} du \frac{{\cal R}(u)}{(1+u)^2}}.
\ee
Since the integral appearing in the above equation is convergent (because ${\cal R}(u) \leq 1$), this result suggests that 
${Y_2^{dyn}}$ is finite when  $\mu < 1$. As furthermore we find that $R_0$ vanishes linearly when $\mu \to 1^-$, we conjecture that: 
\be
Y_2^{dyn} \propto \, 1-\mu \qquad (\mu \to 1)
\ee
which is compatible with the numerics and also comparable with the equilibrium behaviour. The same level of approximation on $Y_3$ also leads to a finite limit $Y_3^{dyn}$, and to a linear temperature behaviour $Y_3^{dyn} \propto \, 1-\mu$ ($\mu \to 1$), so that one can reasonably guess that this linear dependence is valid for all $k>1$.

\begin{figure}
\centerline{
\epsfxsize = 8.5cm
\epsfbox{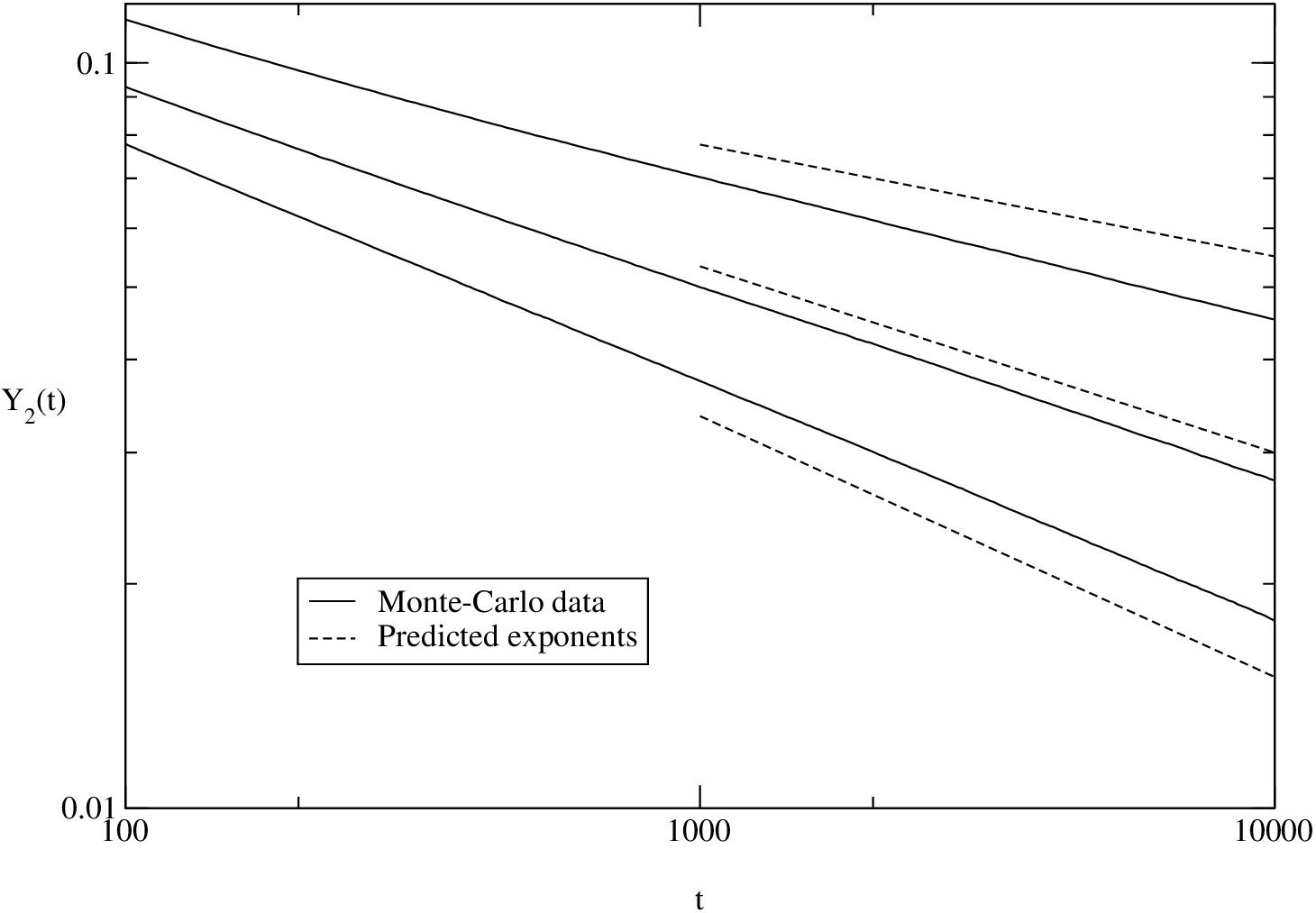}
}
\vskip 0.5 cm
\caption{\sl Plot of $Y_2(t)$ for $\mu=1.3$, $1.5$ and $1.7$ (full lines, from top to bottom), showing a power law decay compatible with the predicted behaviour $t^{(1-\mu)/2}$ (dashed lines). Note that the sub-leading corrections become stronger when $\mu$ is close either to $\mu=1$ or to $\mu=2$.}
\label{Y2_15}
\end{figure}

The case $\mu>1$ has also been studied; we find for $1<\mu<2$ the new predictions:
\bea
\hat{R}(s,s) \sim \frac{1}{s^{\frac{5-\mu}{2}}}\quad(1<\mu<2) \\ \nonumber
\hat{R}(s,s) \sim \frac{1}{s^{\frac{3}{2}}}\quad(\mu > 2)
\eea
which predicts that $Y_2(t)$ tends to zero as $t^{(1-\mu)/2}$ when $1 < \mu < 2$, and as
$t^{-1/2}$ whenever $\mu > 2$. The last result is indeed expected: when the second moment of $\psi(\tau)$
exists, diffusion is normal with no anomalous corrections. The probability that two particles 
starting at time $0$ at the same site happen to be again on the same site at time $t$ decays as: $\xi(t)\times 1/\xi(t)^2 \propto 1/\sqrt{t}$. The result for $1 < \mu < 2$ has 
been checked numerically for $\mu=1.3$, $1.5$ and $1.7$ (see Fig.~\ref{Y2_15}).

\subsection{Partial equilibrium in a finite region} \label{PEC}

We have seen that the dynamical participation ratio never reaches the static
equilibrium value. Can one however isolate a region of space, of size $\ell(t)$ 
possibly much smaller than $\xi(t)$, such that inside that region equilibrium is reached ?
In order to test this idea, one can define a spatially restricted participation ratio $Y_k(\ell,t)$ in the following way as:
\be
Y_k(\ell,t) = \sum_{|i| \leq \ell} \tilde{P}_i(t)^k,
\ee
where $\tilde{P}_i(t)$ is the probability that the walk is on site $i$ conditioned to the fact that it is within the interval $[-\ell,\ell]$:
\be
\tilde{P}_i(t) = \frac{P_i(t)}{\sum_{|j| \leq \ell} P_j(t)}
\ee

\begin{figure}
\centerline{
\epsfxsize = 8.5cm
\epsfbox{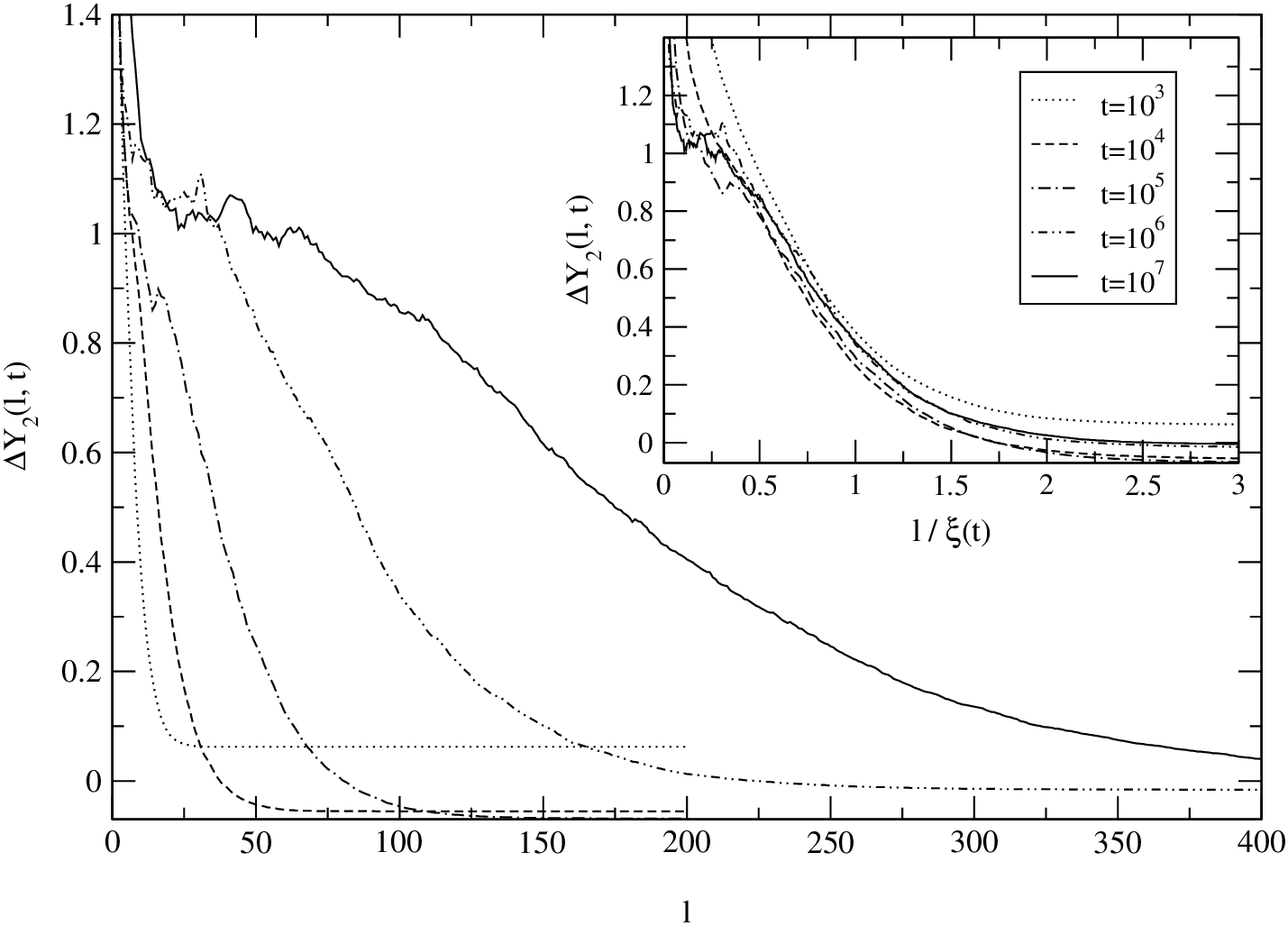}
}
\vskip 0.5 cm
\caption{\sl Plot of $\Delta Y_2(\ell,t)$ versus $\ell$ for $\mu=0.5$ and $t=10^3$, $10^4$, $10^5$, $10^6$ and $10^7$, so that $\xi(t)$ ranges from $10$ to $215$. Inset: the size $\ell$ of the window is rescaled by $\xi(t)$, and the resulting collapse of the curves shows that the equilibration length scale is of the order of a fraction of $\xi(t)$. The strong increase of $\Delta Y_2(\ell,t)$ for small $\ell$ is due to small size effects.}
\label{Y2a}
\end{figure}

Fig.~\ref{Y2a} shows the numerical results for the following rescaled 
quantity:
\be
\Delta Y_2(\ell,t) = \frac{Y_2(\ell,t)-Y_2^{dyn}}{Y_2^{eq}-Y_2^{dyn}},
\ee
such that $\Delta Y_2=1$ for an equilibrated region, and $0$ by construction for
$\ell \gg \xi(t)$.

The results are obtained with $t$ ranging from $10^3$ to $10^7$, and $\mu=0.5$. When $t$ goes to $\infty$ at fixed $\ell$, $Y_2(\ell,t)$ is seen to converge to the corresponding equilibrium value (which depends slightly on $\ell$: Eq.~\ref{Ykeq} is only valid in the limit of large sizes). In the inset we show $\Delta Y_2(\ell,t)$ as
a function of $\ell/\xi(t)$. The collapse is rather good, showing that the size up to
which the system is equilibrated grows as $\xi(t)$, which is thus the only dynamical length scale of this model. That $Y_2(\ell,t)$ is equal to the equilibrium value for 
$\ell < \phi \xi(t)$, where $\phi$ is a small number, means that only the 
`contemporary' processes concerning the largest scale $\xi(t)$ are out of equilibrium.
In a sense, this could have been expected. However, let us emphasize again that a simple description such as Eq.~(\ref{piwithform}), which describes the lack of equilibrium on the scale of $\xi$ through the form factors $g_i(t)$, cannot explain the observed difference
between $Y_2^{eq}$ and $Y_2^{dyn}$.

\section{Generalized equilibrium and half-space excursions}

\subsection{A functional relation between the $Y_k$'s}

\begin{figure}
\centerline{
\epsfxsize = 8.5cm
\epsfbox{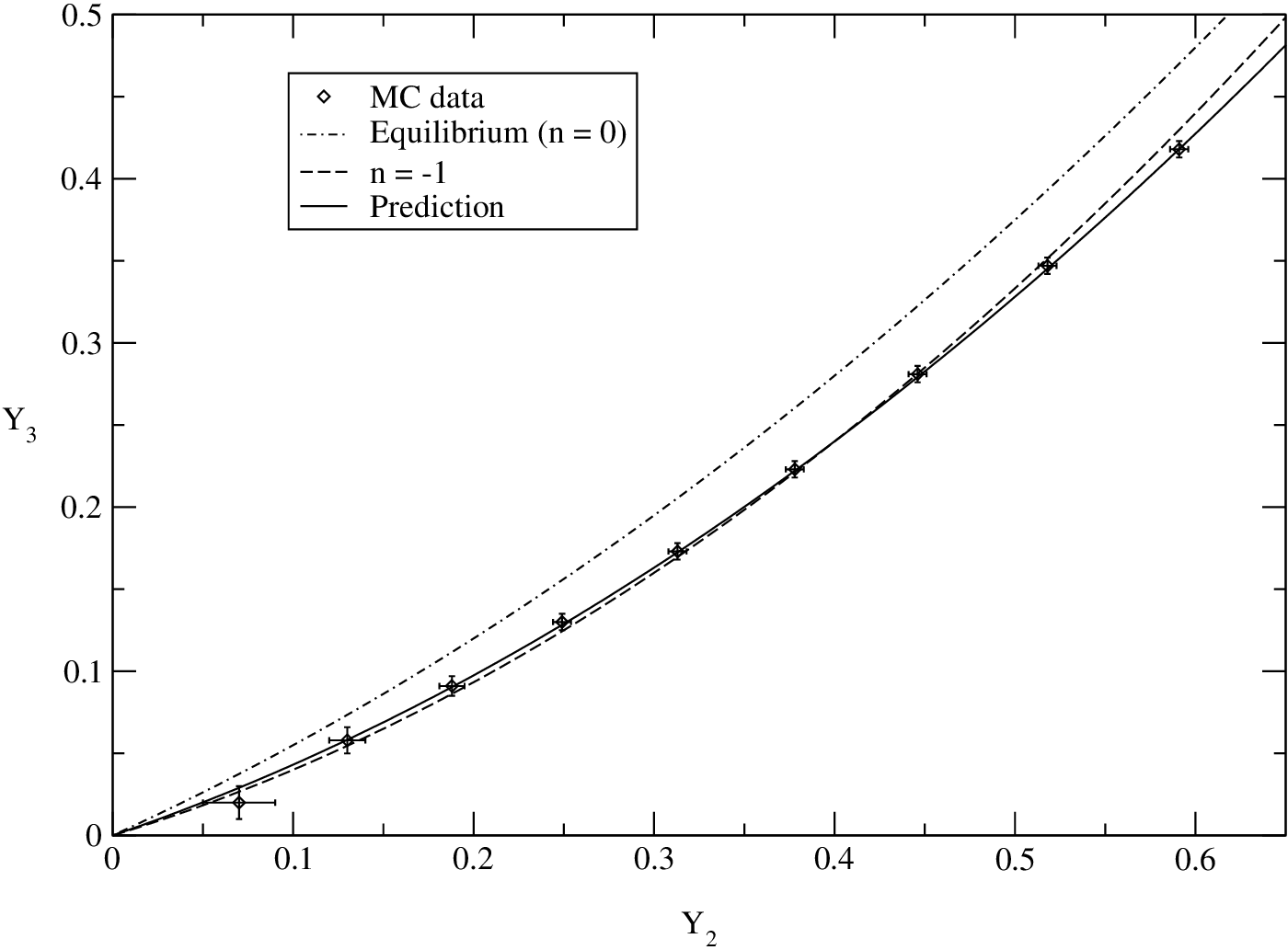}
}
\vskip 0.5 cm
\caption{\sl Plot of $Y_3^{dyn}(\mu)$ versus $Y_2^{dyn}(\mu)$, parameterized by $\mu$. The equilibrium relation, corresponding to $n=0$ in the replica language, is shown for comparison (dotted line). A rather good account of the data is obtained using $n=-1$ (dashed line). However, as we show below, a better description is obtained using a mixture of $n=-2$ cases, see Eq.~(\ref{ykstar}) (full line).}
\label{Y3Y2}
\end{figure}

We have seen that the dynamical participation ratios do not take their
equilibrium value. Could one redefine an effective temperature $\tilde \mu$ 
such that all $Y_k$ can be expressed as equilibrium values with this effective
temperature? In order to test this idea, one can eliminate $\mu$ from the
relation (\ref{Ykeq}), and re-express all $Y_k$ as a function of $Y_2$. One finds:
\be
Y_k = f_k(Y_2) = \frac{\Gamma(k-1+Y_2)}{\Gamma(k)\, \Gamma(Y_2)}
\ee
We have plotted in Fig.~\ref{Y3Y2} $Y_3^{dyn}$ versus $Y_2^{dyn}$ for several values of $\mu$. It appears clearly that this relation is different from the equilibrium one, shown for comparison. This rules out the possibility of defining a meaningful temperature from $Y_2^{dyn}$. 

The above relation between the $Y_k$ is known to be incorrect in other models, such as in the random map for example (see \cite{Derrida}). Inspired from the replica method, one can formally generalize Eq.~(\ref{Ykeq}) to:
\be
Y_{k,n} = \frac{\Gamma(1-n)\,\Gamma(k-\mu)}{\Gamma(k-n)\,\Gamma(1-\mu)},
\ee
where $n$ is the number of replicas (that must be set to $n=0$ to recover Eq.~(\ref{Ykeq}),
see \cite{Derrida,BouMe97}). We can then express $Y_{k,n}$ as a function of $Y_{2,n}$ for arbitrary $n$. We found that the value $n=-1$ gives a reasonable account of the data 
for all $\mu$ values. Interestingly, this value $n=-1$ was found to describe exactly the `area preserving random map model' considered in \cite{Derrida} (where other models, corresponding to different negative values of $n$,
where also studied). However, as we discuss now, a more precise investigation of the problem shows that $n=-1$ does not fully describe our results.

\subsection{The dynamical distribution of weights}

\begin{figure}
\centerline{
\epsfxsize = 8.5cm
\epsfbox{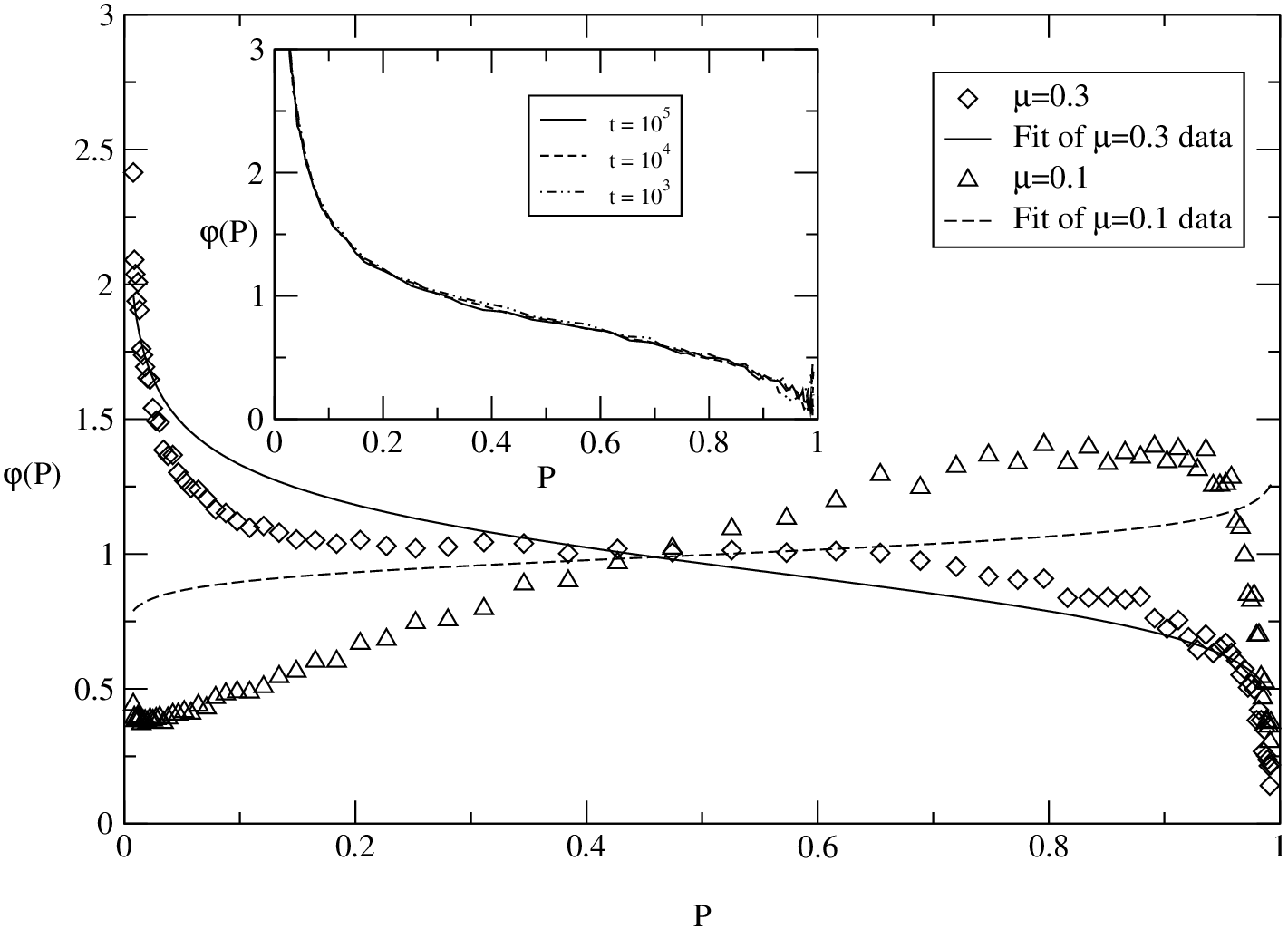}
}
\vskip 0.5 cm
\caption{\sl Fit of $\varphi(P)$ with the $n=-1$ ansatz -- Eq.~\ref{phi_n1} -- using $\tilde{\mu}$ as a free parameter, for $\mu=0.3$ and $\mu=0.1$. The fit is worse and worse as $\mu$ is decreased. Note that for higher values of $\mu$, the fits are better. Inset: $\varphi(P,t)$ is plotted for different times $t=10^3$ (full line), $10^4$ (dashed line) and $10^5$ (dotted line) in order to evidence the convergence towards an out of equilibrium distribution $\varphi_{dyn}(P)$ ($\mu=0.5$).}
\label{conv_phit}
\end{figure}

Instead of studying all the different $Y_k$'s, one can analyse directly the 
time evolution of the distribution of weights, $\varphi(P,t)$, defined as 
(see also Eq.~(\ref{varphi-def})):
\be
\varphi (P,t) = \la\, \sum_i P\, \delta\left(P-P_i(t)\right) \,\rat \qquad 0<P<1
\ee 
For long times in an infinite system, this distribution is expected to reach a
stationary distribution $\varphi_{dyn}(P)$. The inset of Fig.~\ref{conv_phit} shows $\varphi (P,t)$ for three successive (large) times: $t=10^3$, $10^4$ and $10^5$. All three curves collapse rather well, at least not to close to the `edges' $P=0$ and $P=1$,
showing that we are close to the asymptotic distribution. (However, since the 
$Y_k$'s are sensitive to the region around $P=1$, the discrepancies at the edges explain why these moments converge more slowly).

One can define a generalized distribution $\varphi_{n,\tilde{\mu}}(P)$ as the one generating the $Y_{k,n}(\tilde{\mu})$, which leads to the following beta distribution:
\be \label{phi_n1}
\varphi_{n,\tilde{\mu}}(P)= \frac{\Gamma(1-n)}{\Gamma(1-\tilde{\mu})\,\Gamma(\tilde{\mu}-n)} P^{-\tilde{\mu}} (1-P)^{\tilde{\mu}-n-1}
\ee
Following the same line of thought as in the previous subsection, we want to check more precisely if the data are compatible with $n=-1$.

Taking $\tilde{\mu}$ as a free parameter, one can try to fit the numerical data. For $\mu \geq 0.5$ the fits obtained are correct (data not shown). On the contrary, for $\mu < 0.5$ the best fits appear to be quite unsatisfactory, in particular for $\mu=0.1$ (see Fig.~\ref{conv_phit}), showing that the $n=-1$ ansatz does not fully account for the numerical data. This is due to the fact that $\varphi_{n,\tilde{\mu}}(P)$ is a monotonous function whatever the value of $\tilde{\mu}$, whereas $\varphi_{dyn}(P)$ becomes non monotonous for low values of $\mu$.

\subsection{A simple analytical argument in the limit $\mu \to 0$}

Although the out of equilibrium localization problem seems to be hard to tackle at finite temperature, a simple argument can be given in the limit $\mu \to 0$. This argument 
accounts for the non trivial limits $Y_2^{dyn}$ and $Y_3^{dyn}$, which were found to be less than $1$ for $\mu \to 0$ (see Fig.~\ref{Ykmu}). If $\mu$ is very small, then the largest trapping times accessible after a given time $t$ are strongly separated from 
each other. One can for example show that the distribution of the ratio $R$ of the second largest time over the largest is $p(R)=\mu R^{\mu-1}$, which tends to $\delta(R)$ when
$\mu \to 0$. Therefore, in this limit, one can assume that the time elapsed before 
finding the deepest trap $i_0$ occupied at time $t$ is negligible compared to the time spent in $i_0$.\footnote{The time spent on site $i_0$ is actually much greater than the trapping time $\tau_{i_0}$, since the particle comes back to it a large number of times before finding a deeper trap (see section \ref{sect-correl}).}

\begin{figure}
\centerline{
\epsfxsize = 8.5cm
\epsfbox{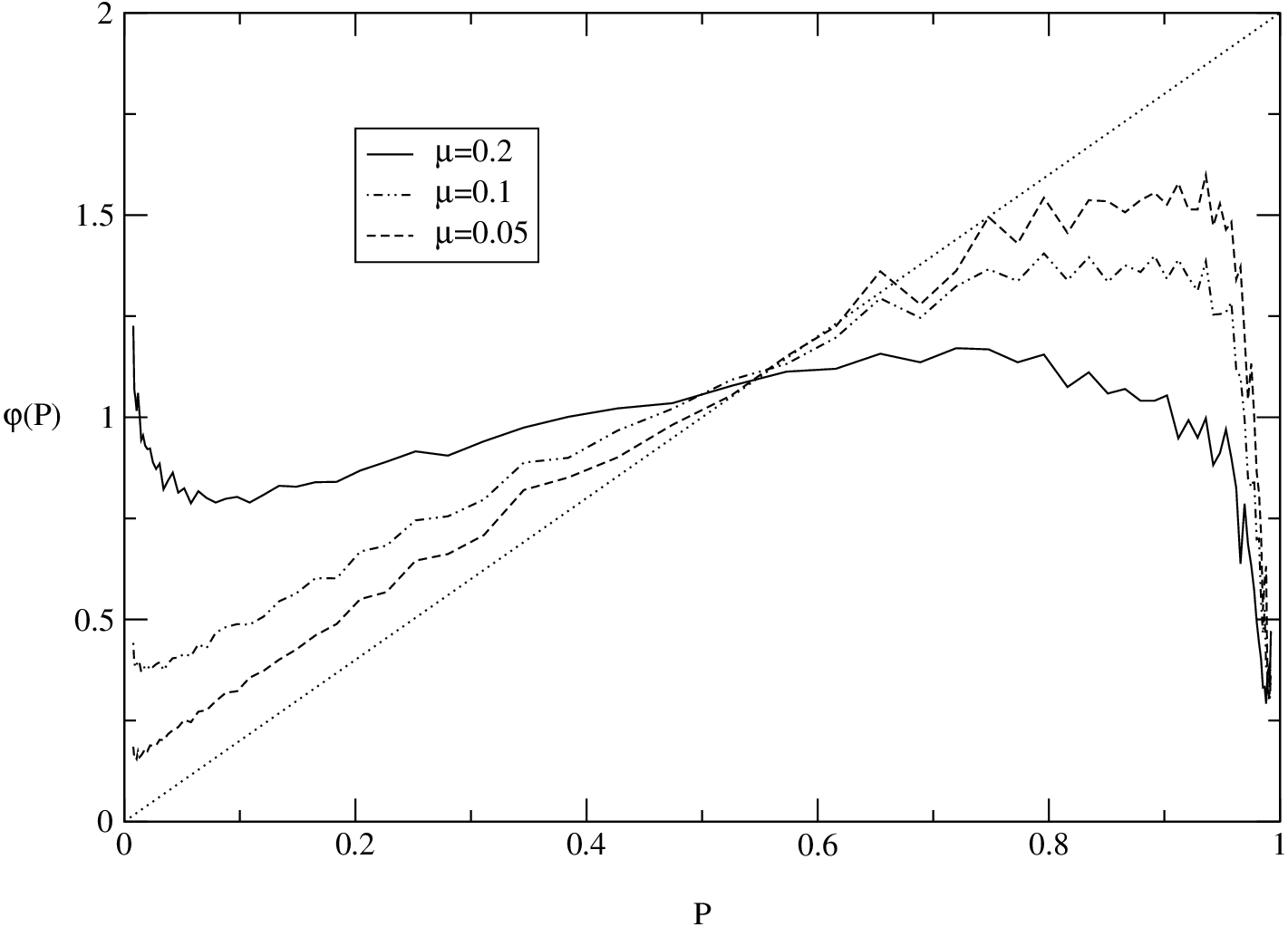}
}
\vskip 0.5 cm
\caption{\sl $\varphi(P)$ for several small values of $\mu$: $\mu=0.2$ (full line), $0.1$ (dot-dashed line) and $0.05$ (dashed line). The curves seem to converge towards the asymptotic distribution $\varphi_0(P)$ (dotted line) for $\mu \to 0$, but the convergence looks quite slow. Note that $P=0$ and $P=1$ are presumably singular points in this convergence process.}
\label{phimu0}
\end{figure}

So the problem becomes equivalent to that of a random walk with no random potential, but with two absorbing boundaries (i.e. the traps with trapping time $>t$ to the right and to the left of the initial site) at random positions. If these absorbing sites are at distances, 
respectively $x_r$ and $x_l$ from the initial position of the walk, then the probability to be absorbed by (say) the left boundary is $p_l=x_r/(x_l+x_r)$. Since the initial
site can be anywhere between these two sites with equal probability, one finds that $p_l$ is a random variable uniformly distributed over $[0,1]$. Coming back to the trap model, it means that only two sites can be occupied, and the corresponding occupation rates are uniform random variables. More precisely $\varphi(P)$ can be written as:
\be
\varphi_0(P) = P \int_0^1 dp_l \left[\delta(p_l-P) + \delta(1-p_l -P) \right] = 2P 
\ee
which leads for $Y_k$ to:
\be
Y_k^0 =  \int_0^1 P^{k-1} \varphi_0(P) dP = \frac{2}{k+1}
\ee
For the particular cases $k=2$ and $k=3$ presented above, this gives $Y_2^0=\frac{2}{3}$ and $Y_3^0=\frac{1}{2}$, which agrees rather well with what can be extrapolated from the numerical data on Fig.~\ref{Ykmu}. Moreover, Fig.~\ref{phimu0} confirms that $\varphi(P)$ converges towards $\varphi_0(P)$ when $\mu \to 0$, although rather slowly. Note also that $\varphi_0(P)$ can be
written in the general form Eq.~(\ref{phi_n1}) given in the previous subsection, for the 
special choice $n=-2$ and $\tilde{\mu}=-1$.

\subsection{Generalization of the argument to finite temperature} \label{Sect-gen-arg}

\begin{figure}
\centerline{
\epsfxsize = 8.5cm
\epsfbox{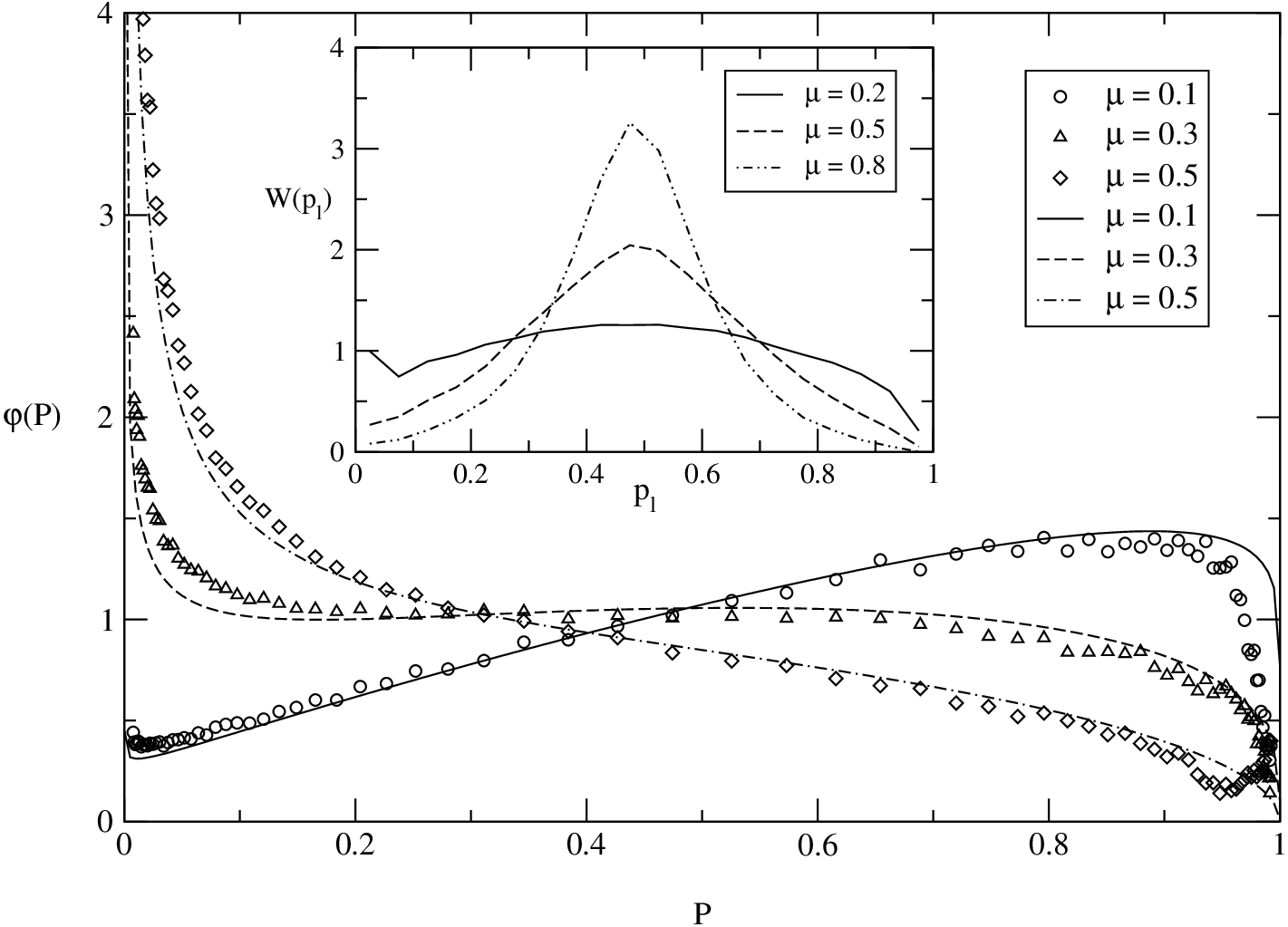}
}
\vskip 0.5 cm
\caption{\sl Comparison between $\varphi(P)$ obtained by numerical simulations (symbols) and $\varphi^*(P)$ given by the argument developed in the text (lines). The agreement appears to be quite good, bearing in mind that no free parameter is used. Inset: distribution $W(p_l)$ of the probability weight $p_l$ carried by one half space, for different $\mu$. This distribution appears to be non uniform (except for $\mu \to 0$), at variance with the hypothesis underlying Eq.~(\ref{varphi}).}
\label{Gen-arg}
\end{figure}

The above argument can be reinterpreted in the following way. In equilibrium, the zero temperature limit means that a single site dominates and contains all the probability 
weight. This is why $Y_k^{eq} \to 1$ when $\mu \to 0$. On the other hand, in one 
dimension, the time needed to explore an interval of size $L$ is $L^{(1+\mu)/\mu}$, 
which grows much faster than the time to exit the deepest traps ($\sim L^{1/\mu}$) found in the interval.
Therefore, if a deep trap is encountered in -- say -- the left region of the
line, there is a substantial probability that the particle will not have time to
explore the right region and equilibrate with a trap of comparable depth. 
This is the essence of the above argument: at zero temperature, the fraction $p_l$ of
the weight captured by the left trap is uniform between zero and one, independently
of the relative depth of the two traps. A simple way to generalize this argument for finite temperature is to assume that each half-space is independently equilibrated, and carries a total weight uniformly distributed between zero and one, as in the zero temperature limit. (Actually, as noticed above, we only need to assume that in each half-space the probability
distribution has the form given by Eq.(\ref{piwithform}) with arbitrary factors $g_i$: 
this does not affect the asymptotic shape of $\varphi(P)=\varphi_{eq}(P)$.)
Denoting by $\varphi(P,p_l)$ the distribution restricted to the left half-space, normalized to $p_l$, one has for $0<P<p_l$:
\be
\varphi(P,p_l) = \frac{p_l}{\Gamma(1-\mu)\, \Gamma(\mu)}\, P^{-\mu}\, (p_l-P)^{\mu-1}
\ee
Averaging over $p_l$ with a uniform weight, and taking into account the right half-space, leads to the following prediction for $\varphi(P)$:
\bea\label{varphi}
\varphi(P) &\simeq& \varphi^*(P) = 2 \int_0^1 dp_l\, \varphi(P,p_l)\, \theta(p_l-P)\\
\varphi^*(P) &=& \frac{2}{\Gamma(1-\mu)\,\Gamma(1+\mu)}\,P^{1-\mu} (1-P)^{\mu}\\
&+& \frac{2\mu}{\Gamma(1-\mu)\,\Gamma(2+\mu)}\,P^{-\mu} (1-P)^{1+\mu} \nonumber
\eea
Although we do not have any interpretation for this, one can notice that $\varphi^*$ can be written as a superposition of distributions $\varphi_{-2,\tilde{\mu}}$, with two different values of $\tilde{\mu}$:
\be
\varphi^*(P) = (1-\mu)\, \varphi_{-2,\mu-1}(P) + \mu\, \varphi_{-2,\mu}(P)
\ee
with $\varphi_{n,\tilde{\mu}}$ defined in Eq.~(\ref{phi_n1}), and in agreement with what was found for $\mu \to 0$. This prediction is compared with the numerics in Fig.~\ref{Gen-arg}, for several values of $\mu$ ($\mu=0.1$, $0.3$, $0.5$). The agreement is rather good; note that no fitting 
parameter is used here. We note that the participation ratios $Y_k$ obtained using $\varphi^*(P)$
are simply proportional to the equilibrium values:
\be\label{ykstar}
Y_k^* = \frac{2}{k+1} Y_k^{eq}.
\ee 
As shown in Fig.~\ref{Ykmu}, this relation accounts quite well (but not exactly) for the data.
However, the distribution $W(p_l)$ of the weight $p_l=1-p_r$ carried by 
one half-space is not found to be uniform as we assumed, except for $\mu \to 0$
-- see the inset of Fig.~\ref{Gen-arg}. Surprisingly, if one redo the
above computation with a humped shaped distribution $W(p_l)=A [p_l (1-p_l)]^{\sigma}$, the resulting $\varphi^*(P)$ does not fit 
the data as well as the above form, which corresponds to $\sigma=0$.

\subsection{Discussion}

The physical picture emerging from the last sections is the following: for the purpose of understanding 
global localization quantities, one can reasonably consider that space is split into two half lines, and that each one behaves as if it was independently  equilibrated, but out of equilibrium with respect to the other. On the other hand, we have seen in section \ref{PEC} that length scales much smaller than $\xi$ could be considered as equilibrated, and that departure 
from equilibrium comes from the largest length scale $|x| \gtrsim \xi$. The equilibration of the small scales cannot be accounted for by the previous argument. All these observations suggest, in order to get a consistent picture, that space may be actually divided into three regions, an equilibrated domain centered on the origin and two quasi equilibrated regions on each side, and that each part of space is not equilibrated with the others.

An artificial remedy
to this lack of equilibration between the different regions is to allow the particle to make 
long jumps. We have therefore added `links' between the sites $x$ and $-x$, such that the
probability to hop directly from $x$ and $-x$ decays as $x^{-\rho}$. When $\rho$ is large enough,
the dynamical participation ratio $Y_2(t)$ is a decreasing function of time, and seems
to converge to $Y_2^{dyn}$. When $\rho < \rho_c$ on the other hand, $Y_2(t)$ is seen to reach a
minimum and to increase back towards the equilibrium value $Y_2^{eq}$. We have however not
checked in details whether $Y_2(t)$ indeed converges towards $Y_2^{eq}$ for all $\rho < \rho_c$, but only wanted
to illustrate that the difference between $Y_2^{dyn}$ and $Y_2^{eq}$ is due to the scarcity 
of the links between the different sites for the one dimensional lattice.

\section{Correlation functions, aging and sub-aging} \label{sect-correl}

\subsection{Motivation}

Let us now turn to different correlation functions that one can define in order
to probe the peculiar {\it aging} properties of this model. Since the largest 
encountered trapping time during $t_w$ scales as $t_w^\nu$ with $\nu=\frac{1}{1+\mu} < 1$, one
would naively expect that two time correlation functions vary on a time scale $\sim t_w^\nu$.
This would correspond to `sub-aging' behaviour, where the effective relaxation time grows
less rapidly than $t_w$ itself. 

This is indeed the case for the probability $\Pi(t_w+t, t_w)$ of not having jumped at all
between time $t_w$ and $t_w+t$. This correlation function was computed numerically in 
\cite{Maass}, and was found to scale very accurately as: $\Pi(t_w+t, t_w)=\pi(t/t_w^\nu)$. 
The shape of the scaling function was
compared to the prediction of an approximate calculation where one assumes 
`partial equilibrium', i.e. that the probability to find the particle in a trap of 
depth $\tau$ after time $t_w$ is equal to the equilibrium probability 
within a region of size $\xi(t_w)$. This approximation predicts a power law 
behaviour for $\pi(s)$ both for small and large $s$, with exponents that agree 
with their numerical determination. The detailed shape of $\pi(s)$ however departs from 
the numerical results, which is expected. The success of the partial equilibrium assumption 
here is due to the fact that $\Pi(t_w+t, t_w)$ only depends on the average probability to occupy a site, 
and not on higher order correlations such as needed to compute the participation ratios $Y_k$.

Perhaps surprisingly, different correlation functions may exhibit a completely different aging
behaviour. Consider the probability $C(t_w+t, t_w)$ that the particle occupies the same site 
at time $t_w+t$ and at time $t$. Obviously, $C(t_w+t, t_w) \geq \Pi(t_w+t, t_w)$. 
But in this case, it was shown rigorously in \cite{Isopi,BenArous2} that 
$C(t_w+t, t_w)$ scales as a function of $t/t_w$, and {\it not} as $t/t_w^\nu$. 
This means that even if the particle 
has almost certainly jumped away from its starting point after a time $t_w^\nu \ll t_w$, it has 
returned there even after a time of order $t_w$ so as to make $C(2t_w, t_w)=O(1)$, whereas $\Pi(2t_w,t_w) \to 0$. 
This difference is not intuitive a priori, in particular because one knows that once the particle has
left its initial trap after a time $\sim t_w^\nu$, it takes on average an infinite time to get
back there since the walk is one dimensional. But if $C(t_w+t, t_w)$ is to decay on the scale $t_w$, it means 
that the probability not to find the particle on its starting point after 
a time $t$ much greater than $t_w^\nu$ but much less than $t_w$ must tend to zero when $t_w \to 
\infty$. The fact that the particle jumps back and forth 
a large number of times between $t_w^\nu$ and $t_w$ could thus a priori lead to an 
interesting behaviour of $C(t_w+t, t_w)$ in the short time
regime $t/t_w \ll 1$ (which was not investigated in \cite{Isopi}). For example, one could 
find, as in \cite{Maass}, different `time domains' 
$t \sim t_w^{\nu_1}$, $t \sim t_w^{\nu_2}$, etc., where the correlation function has a 
different analytic behaviour. This is the issue that we discuss below.

\subsection{Analytical arguments}

The difference of scaling between $\Pi$ and $C$ can be qualitatively understood as follows:
for $C(t_w+t,t_w)$ to decay to zero, one has to wait until the region probed by the particle
at time $t_w+t$ is much larger than the initial region where it was located, i.e.~a time $t$
such that $\xi(t_w+t) \gg \xi(t_w)$. But since $\xi(t_w) \sim t_w^{\frac{\mu}{1+\mu}}$, 
the time
needed for $C$ to decay to zero is necessarily of order $t_w$. (Note that this argument does not
hold for $\Pi(t_w+t,t_w)$, which only requires the particle to hop once out of its initial trap.)

In the same spirit as Ref.~\cite{Maass} for $\Pi(t_w+t,t_w)$, but using a slightly different method, one can try to give an approximate calculation of $C(t_w+t,t_w)$. The first step is to introduce the dynamical distribution of trapping times $p(\tau,t_w)$, assumed to behave as:
\be
p(\tau,t_w) \simeq \frac{1}{t_w^{\nu}} \phi \left(\frac{\tau}{t_w^{\nu}}\right).
\ee
This encodes the fact that typical trapping times are of order $t_w^\nu$. If one assumes that short time scales ($\tau \ll t_w^{\nu}$) are equilibrated, whereas large ones ($\tau \gg t_w^{\nu}$) are still distributed according to the a priori distribution (this can be rigorously proved in the fully connected trap model), one obtains the following asymptotic behaviour for $\phi(z)$:
\bea \label{scal-p-tau}
\phi(z) &\simeq& \frac{\gamma_0}{z^{\mu}} \qquad z \to 0\\
\phi(z) &\simeq& \frac{\gamma_{\infty}}{z^{1+\mu}} \qquad z \to +\infty
\eea
Using the relation:
\be
\Pi(t_w+t,t_w) = \int_1^{\infty} d\tau\, p(\tau,t_w)\, e^{-t/\tau}
\ee
one can easily deduce from Eq.~(\ref{scal-p-tau}) the short and late time behaviour of $\Pi(t_w+t,t_w)$:
\bea \label{eqn_cor_pi}
\Pi(t_w+t,t_w) &\simeq& 1 - \frac{\gamma_0}{1-\mu} \Gamma(\mu)\, \left(\frac{t}{t_w^{\nu}}\right)^{1-\mu} \quad t \ll t_w^{\nu} \\
\Pi(t_w+t,t_w) &\simeq& \gamma_{\infty} \Gamma(\mu)\, \left(\frac{t}{t_w^{\nu}}\right)^{-\mu} \qquad \qquad t \gg t_w^{\nu}
\eea
in agreement with the results of Ref.~\cite{Maass}, and with the numerics (see below).

Turning now to $C(t_w+t,t_w)$, one has to take into account the fact that when a particle leaves its trap, it will come back a large number of times before really escaping. We thus propose the following approximation: a particle will be considered to have truly left its 
initial trap if it has encountered a deeper trap during its excursion out of the original trap. Given a trapping time $\tau$, the probability that $\tau' > \tau$ is given by:
\be
P(\tau'>\tau) = \int_{\tau}^{\infty} \frac{\mu d\tau'}{\tau'^{1+\mu}} = \frac{1}{\tau^{\mu}}
\ee
So the probability $\tilde{p}(\ell,\tau)$ that the first trap encountered with a trapping time larger than $\tau$ is found at a distance $\ell$ is:
\be
\tilde{p}(\ell,\tau) = P(\tau'>\tau) \left[ P(\tau'<\tau) \right]^{\ell-1} = \frac{1}{\tau^{\mu}} \left( 1-\frac{1}{\tau^{\mu}}\right)^{\ell-1}
\ee
For large $\tau$'s, one has:
\be
\tilde{p}(\ell,\tau) \approx \frac{1}{\tau^{\mu}}\, e^{-\ell/\tau^{\mu}}
\ee
Note that we only give here a scaling argument, and that corrections coming from the fact that there is a deeper trap on both sides are neglected. Conditioned to the fact that the deeper trap is situated at a distance $\ell$, the particle has a probability $\frac{1}{\ell}$
to reach it, once it has jumped out of its initial trap. Therefore, the escape rate can be written as:
\be
w(\tau,\ell) = \frac{1}{\tau \ell}
\ee
The correlation function $C(t_w+t,t_w)$ is then given by:
\be
C(t_w+t,t_w) \approx \int_1^{\infty} d\tau\, p(\tau,t_w) \int_1^{\infty} d\ell\, \tilde{p}(\ell,\tau) \, e^{-w(\tau,\ell) t}
\ee
After a few changes of variables, and using the scaling relations Eq.~(\ref{scal-p-tau}), one finds the following short time and late time behaviour for $C(t_w+t,t_w)$:
\bea \label{eqn_cor_C}
C(t_w+t,t_w) &\simeq& 1 - c_s \, \left(\frac{t}{t_w}\right)^{\frac{1-\mu}{1+\mu}} \qquad t \ll t_w \\
C(t_w+t,t_w) &\simeq& c_l \, \left(\frac{t}{t_w}\right)^{-\frac{\mu}{1+\mu}} \qquad \quad t \gg t_w
\eea
where the constants $c_s$ and $c_l$ are given by:
\bea
c_s = \frac{\gamma_0}{1-\mu}\, \Gamma\left(\frac{2\mu}{1+\mu}\right)^2 \\
c_l = \frac{\mu \gamma_{\infty}}{(1+\mu)^2}\, \Gamma\left(\frac{\mu}{1+\mu}\right)^2
\eea
These values for the short time and late time singularity exponents, have, to our knowledge, not been reported 
before, although they should in principle be contained in the analysis of \cite{Isopi}.
We now turn to a numerical investigation of these asymptotic predictions.

\subsection{Numerical results and multiple time scales} 

\begin{figure}
\centerline{
\epsfxsize = 8.5cm
\epsfbox{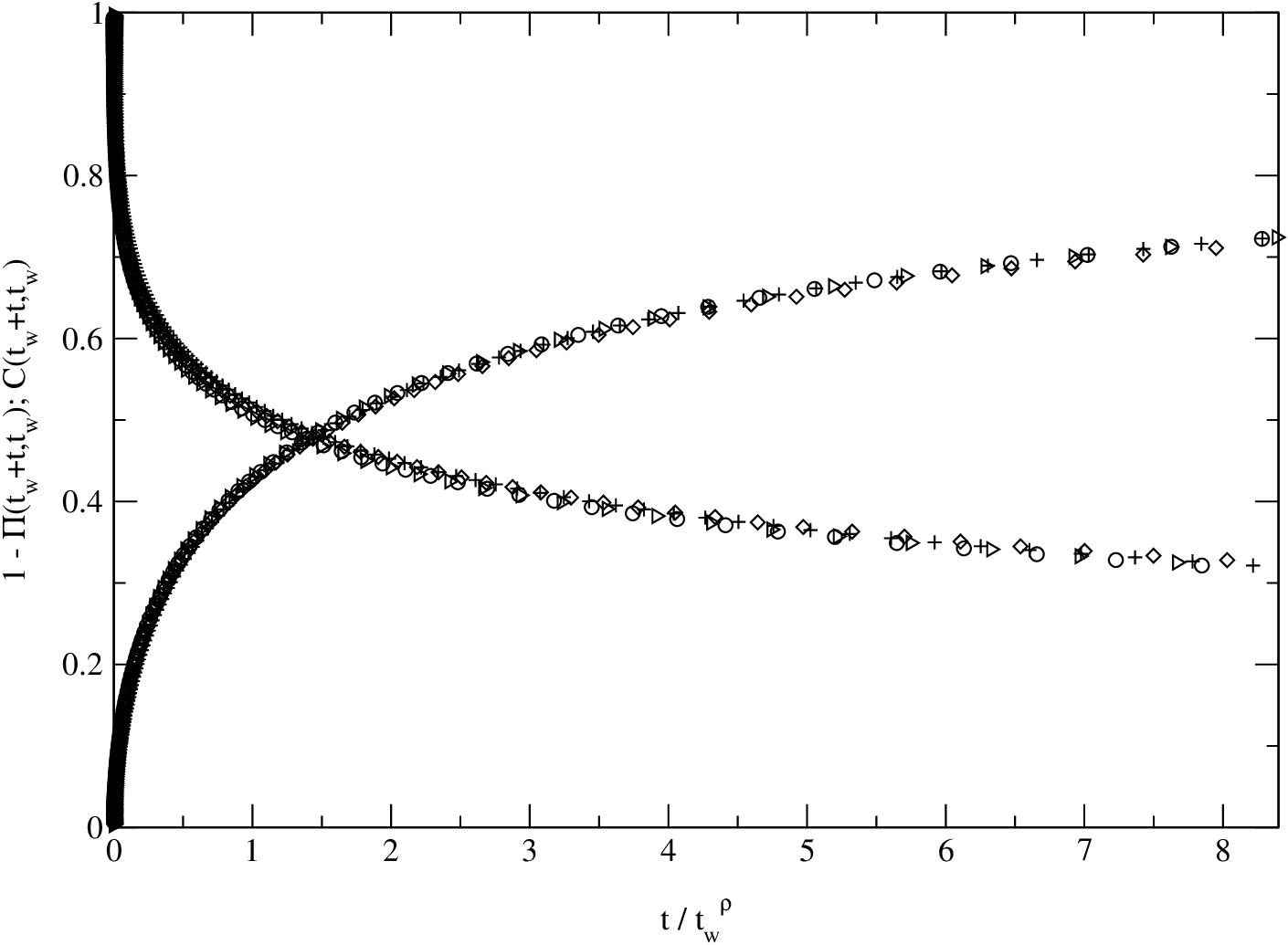}
}
\vskip 0.5 cm
\caption{\sl Plot of $1-\Pi(t_w+t,t_w)$ (increasing curve) versus $t/t_w^{\nu}$ and $C(t_w+t,t_w)$ (decreasing curve) versus $t/t_w$, for $\mu=\frac{1}{2}$, 
and $\nu=\frac{1}{1+\mu}=\frac{2}{3}$.
The scaling relations are very well satisfied, at least in this time window. 
Symbols refer to the same waiting times for the two curves: 
$t_w=10^3$ ($+$), $10^4$ ($\diamond$), $10^5$ ($\circ$) and $10^6$ ($\triangleright$).}
\label{corCpi}
\end{figure}

Fig.~\ref{corCpi} displays $1-\Pi(t_w+t,t_w)$ as a function of $t/t_w^{\frac{1}{1+\mu}}$, 
and $C(t_w+t,t_w)$ as a function of $t/t_w$, for different waiting times 
($t_w=10^3$, $10^4$, $10^5$ and $10^6$) and at temperature $\mu=\frac{1}{2}$. 
The collapse is very satisfactory, confirming the validity of the predicted scaling relations.

Let us analyse in more details the short time behaviour of these correlation functions. 
When plotting $\ln(1-\Pi)$ as a function of $\ln(t/t_w^{\nu})$, the scaling is still quite well obeyed and in good agreement with the theoretical prediction $1-\Pi \sim (t/t_w^{\nu})^{1-\mu}$, obtained in \cite{Maass}, up to small time corrections that vanish only when  $\Gamma_0 t \gg 1$.

On the other hand, a similar plot of $\ln(1-C)$ as a function of $\ln(t/t_w)$ is less convincing, which could be the sign of multiple time regimes (as 
was the case in \cite{Maass}, where similar `non scaling' features actually suggested such 
regimes). A way to investigate this issue is to study the 
function $g(\alpha,t_w)$ defined as:
\be
g(\alpha,t_w) = -\frac{\ln[1-{C}(t_w+t_w^\alpha,t_w)]}{\ln t_w}.
\ee
If this function has a limit $g_{\infty}(\alpha)$ when $t_w \to \infty$, it means that in the time domain where $t \sim t_w^\alpha$, the probability $1-C$ that the particle has escaped from its starting site decays as $t_w^{-g_{\infty}(\alpha)}$ for large $t_w$.
From the $t/t_w$ regime established by \cite{Isopi}, we already know that $g_{\infty}(1)=0$.
If $1-C(t_w+t,t_w)$ behaves as $(t/t_w)^\lambda$ even for $t \sim t_w^\alpha$ with $\alpha < 1$,
then one should observe $g_{\infty}(\alpha)=\lambda(1-\alpha)$. Any departure from a 
linear function $g_{\infty}(\alpha)$ would signal multiple time regimes; 
in particular, for the model considered in \cite{Maass} in $d=1$ where two sub-aging 
exponents $\nu_2 < \nu_1 < 1$ appear, one finds that the function $g_{\infty}(\alpha)$ 
is piecewise linear in the intervals $[0,\nu_2]$ and $[\nu_2,\nu_1]$, with different slopes.
One also finds that $g_{\infty}(\nu_2^-)=g_{\infty}(\nu_2^+)$, and $g_{\infty}(\alpha > \nu_1)=0$.
In this case, the change of slope indicates the presence of a characteristic time scale.
One could imagine more complicated  `multi-scaling' situations where $g_{\infty}(\alpha)$ 
is a non trivial curve.

\begin{figure}
\centerline{
\epsfxsize = 8.5cm
\epsfbox{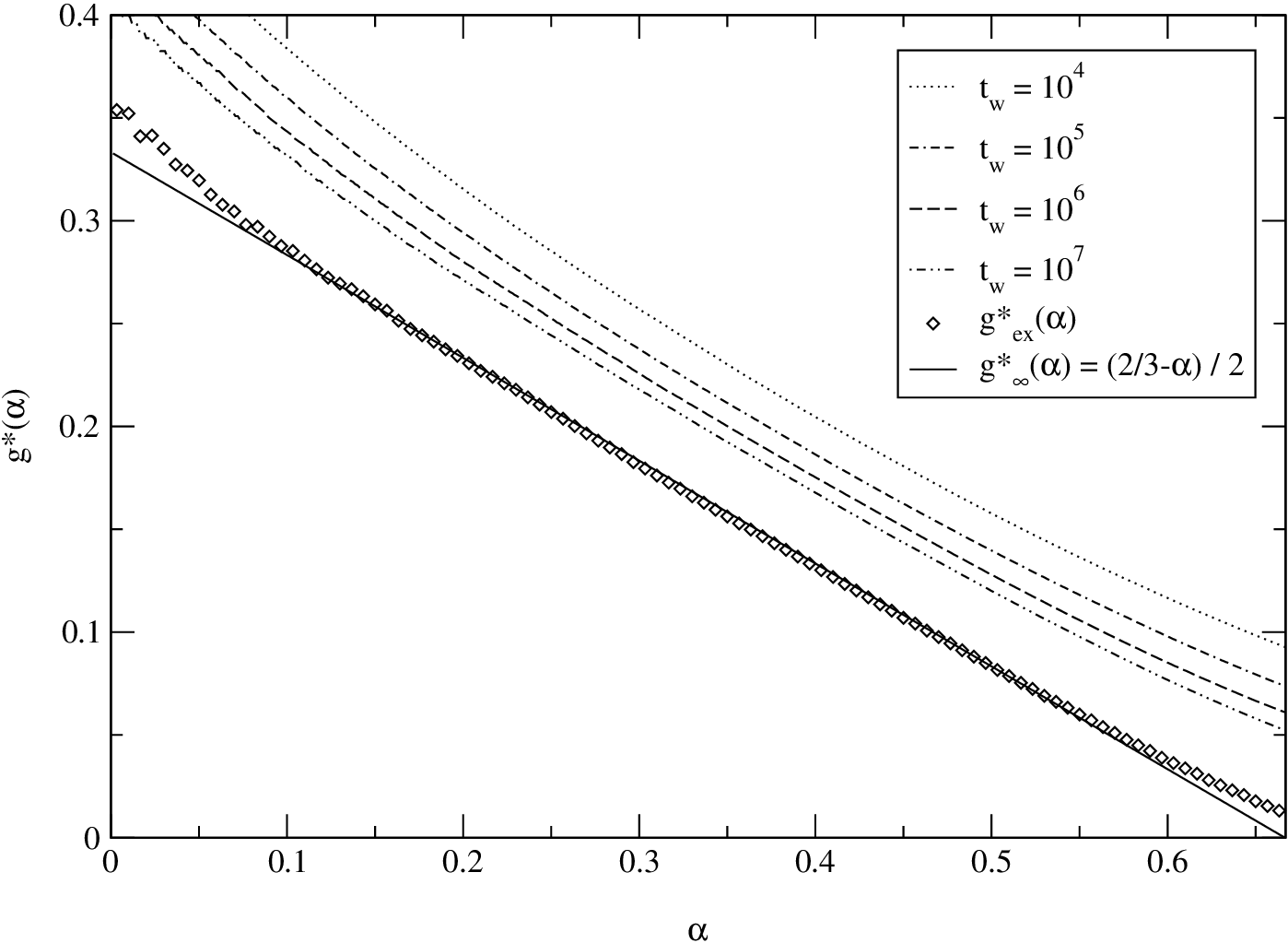}
}
\vskip 0.5 cm
\caption{\sl Function $g^*(\alpha,t_w)$, associated to $\Pi(t_w+t,t_w)$, with $\mu=\frac{1}{2}$ and $t_w = 10^4$, $10^5$, $10^6$ and $10^7$. The function $g^*_{\infty}(\alpha)$ (full line) is expected to be $g^*_{\infty}(\alpha)=\frac{1}{2}(\frac{2}{3}-\alpha)$ (see text). 
The infinite time extrapolation $g^*(\alpha,t_w)$ ($\diamond$) agrees very well with the prediction, with small discrepancies near $\alpha=0$ and $\alpha=\frac{2}{3}$, where finite time effects are stronger.} 
\label{cor_pi}
\end{figure}

We have first tested this procedure on $\Pi(t_w+t,t_w)$, defining in the same way a function $g^*(\alpha,t_w)$ associated to $\Pi$. In this case, a single sub-aging scaling is expected, with $\nu=\frac{1}{1+\mu}$, and $\lambda=1-\mu$, which leads to $g_{\infty}(\alpha) = (1-\mu)(\nu-\alpha)$. The function $g^*(\alpha,t_w)$ is plotted for different values of $t_w$ (namely $t_w=10^4$, $10^5$, $10^6$ and $10^7$) in Fig.~\ref{cor_pi}, for $\mu=\frac{1}{2}$. 
On general grounds, one expects finite time corrections to $g^*_{\infty}(\alpha)$ that decays as 
$1/\ln t_w$ and $1/t_w^{\gamma}$. 
Using these corrections, one can very satisfactorily extrapolate $g^*_\infty(\alpha,t_w)$ to a function 
which is very close to the expected result $g^*_{\infty}(\alpha)=\frac{1}{2}(\frac{2}{3}-\alpha)$ (see Fig.~\ref{cor_pi}). To be more specific, 
we used the following functional form for the extrapolation:
\be
g^*(\alpha,t_w) = g^*_{\infty}(\alpha) + \frac{b}{\ln t_w} + c(\alpha) t^{-\gamma(\alpha)}
\ee
where $g^*_{\infty}(\alpha)$, $c(\alpha)$ and $\gamma(\alpha)$ are fitted for each value of $\alpha$, and $b$ is a fitting coefficient independent of $\alpha$, since the $1/\ln t_w$ correction is expected to come from the prefactor of $(t/t_w^{\nu})^{1-\mu}$ in the short time expansion of the correlation function. Therefore the value of $b$ was fixed from the direct power law fit of the short time regime of $1-\Pi$.

\begin{figure}
\centerline{
\epsfxsize = 8.5cm
\epsfbox{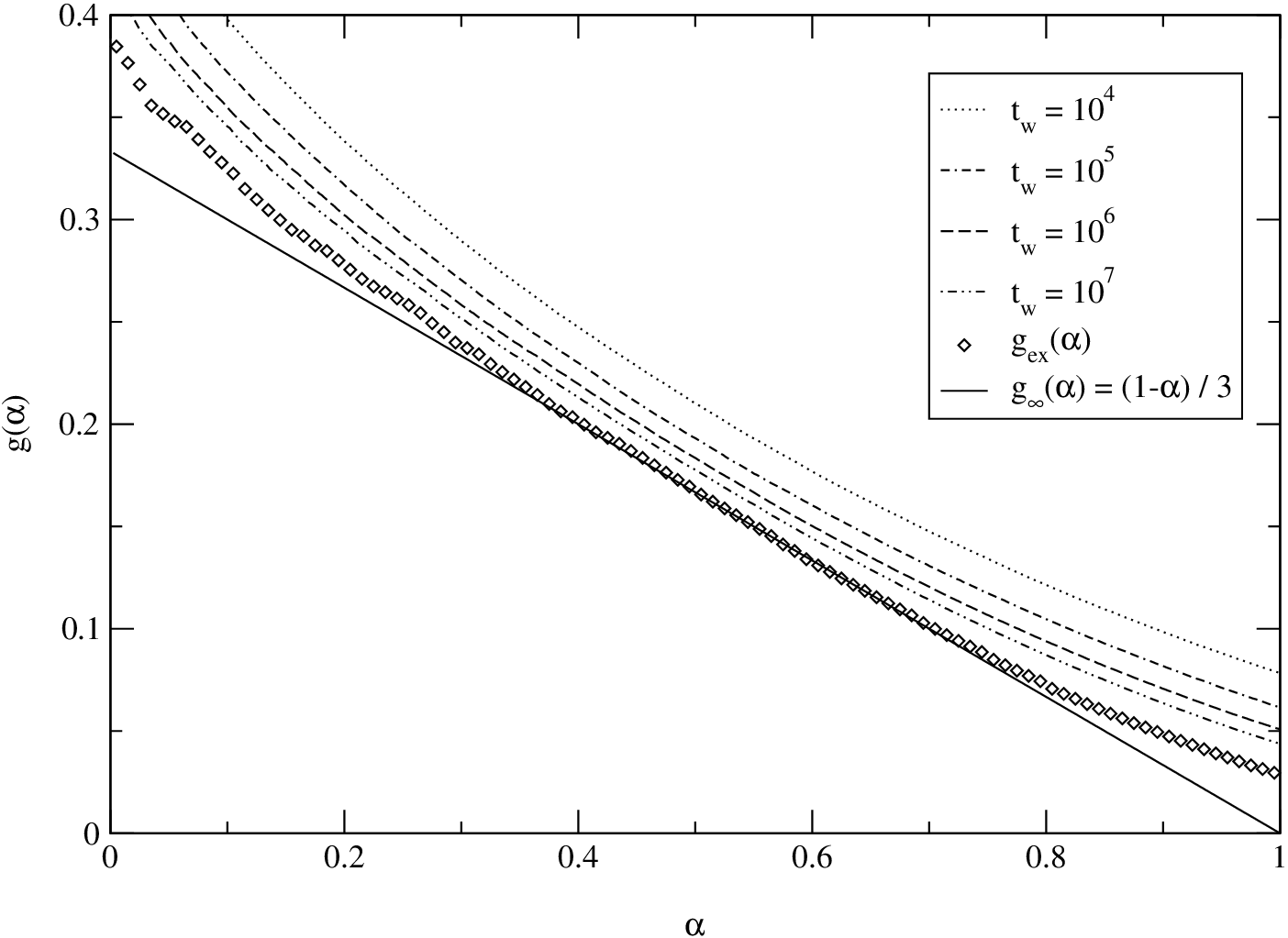}
}
\vskip 0.5 cm
\caption{\sl Function $g(\alpha,t_w)$, associated to $C(t_w+t,t_w)$, for $\mu=\frac{1}{2}$ and with the same waiting times as for $g^*(\alpha,t_w)$. The argument developed in the text -- Eq.~(\ref{eqn_cor_C}) -- predicts $g_{\infty}(\alpha)=\frac{1}{3}(1-\alpha)$ (full line). Although finite time corrections are stronger than in the previous case, the infinite time extrapolation $g_{ex}(\alpha)$ ($\diamond$) agrees well with the prediction, at least for $0.2<\alpha<0.8$.}
\label{cor_C}
\end{figure}

One can now apply the same procedure to $C(t_w+t,t_w)$. 
The results are shown in Fig.~\ref{cor_C}, using the same convention as for 
Fig.~\ref{cor_pi}; $g(\alpha,t_w)$ is represented for the same waiting times as $g^*(\alpha,t_w)$. 
Interestingly, although finite time corrections are strong, the extrapolated 
results are in good agreement with our analytical prediction  
$g_{\infty}(\alpha)=\lambda(1-\alpha)$, with $\lambda=(1-\mu)/(1+\mu)=1/3$, at least when $\alpha \in [0.2,0.8]$. 
This suggests that a unique time regime $t \sim t_w$ is relevant for $C(t_w+t,t_w)$, although we know that the time scale $t_w^\nu \ll t_w$ governs the evolution of $\Pi$. Note that the $1/\ln t_w$ corrections are weaker than in the previous case, and one is almost dominated by power law corrections. This is due to the fact that the prefactor of $(t/t_w)^{\lambda}$ 
in the short time regime happens to be close to $1$ here (and hence the parameter $b$ is small), whereas the prefactor of $(t/t_w^{\nu})^{\lambda}$ was about $0.57$ for $\Pi$.

So, what happens to the particles that have left the initial trap after a short time $t_w^\nu$ and took a very long time to come back? The probability that a particle leaves the trap exactly at $t'$ is $\partial [1-\Pi(t_w,t_w+t')]/\partial t' \sim t'^{-\mu}/t_w^{\nu(1-\mu)}$.
If the sample was not disordered, the probability that it has not returned to the origin after time $t-t'$ decays as $(t-t')^{-1/2}$. Because of the long trapping times, this probability actually decays slower, as $(t-t')^{-\mu/(1+\mu)}$. These particles contribute to $1 - C$, as:
\be
1 - C(t_w+t, t_w) \sim \int_0^t d t' \, \frac{t'^{-\mu}}{t_w^{\nu(1-\mu)}} 
(t-t')^{-\mu/(1+\mu)}.
\ee
Choosing $t = t_w^\alpha$, we find that the contribution of these `early birds' to $1 - C$ is a factor $t_w^{-\alpha \mu^2/(1+\mu)}$ smaller than the contribution computed above, Eq.~(\ref{eqn_cor_C}), and are thus negligible in the large $t_w$ limit.

This simple estimate shows (i) why finite time corrections become large when $\alpha \to 0$,
(ii) that power-law corrections to $g(\alpha,t_w)$, as the one used for our
extrapolation, are indeed expected and finally (iii) justifies why the $t_w^\nu$ time
scale does not appear in $C(t_w+t,t_w)$.

\begin{figure}
\centerline{
\epsfxsize = 8.5cm
\epsfbox{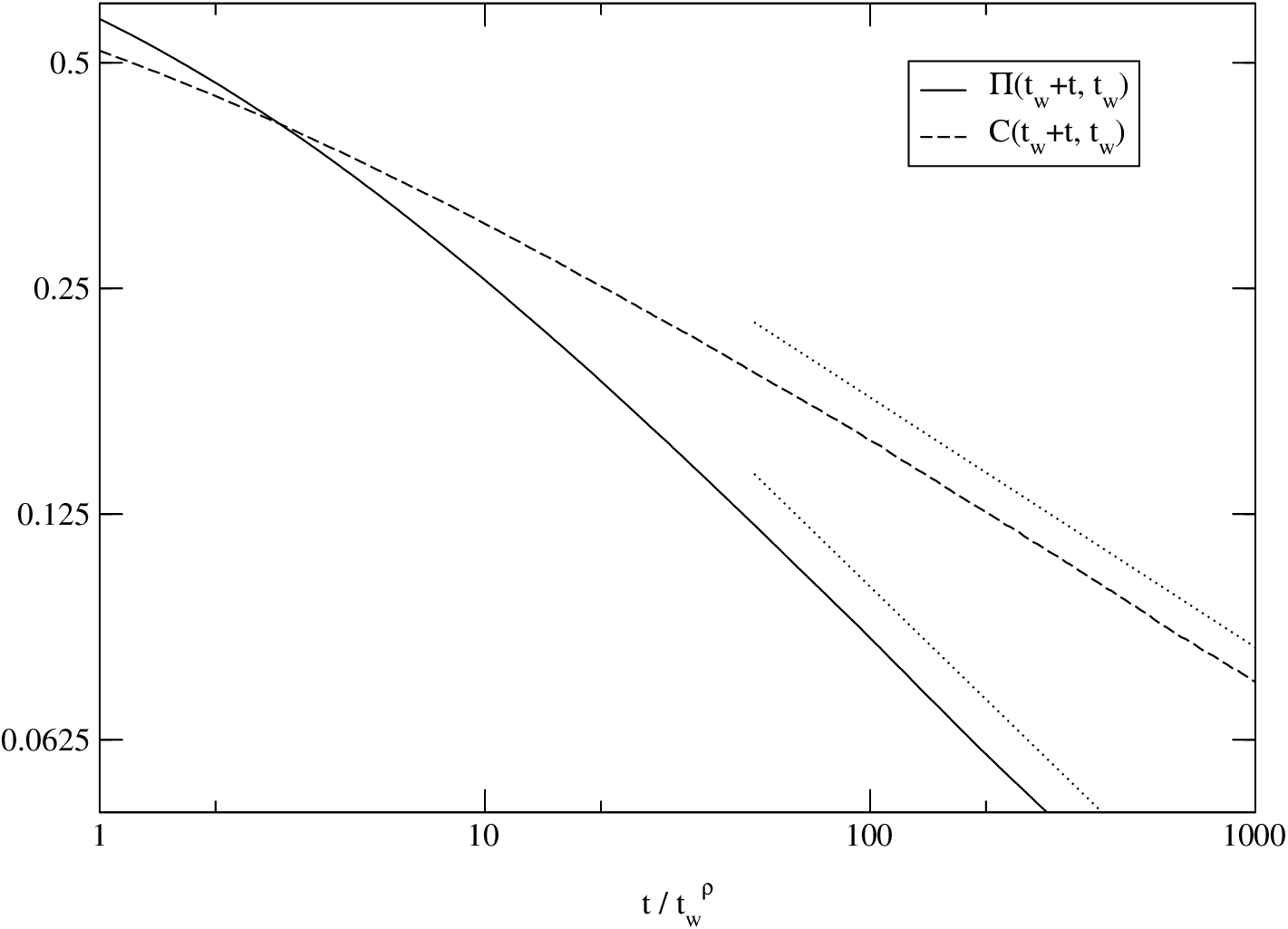}
}
\vskip 0.5 cm
\caption{\sl Plot of the late time behaviour of $\Pi(t_w+t,t_w)$ versus $t/t_w^{\nu}$, and $C(t_w+t,t_w)$ versus $t/t_w$, for $t_w=10^5$ and $10^3$ respectively, and $\mu=0.5$. Both correlation functions exhibit a power law behaviour at large time, and the exponents agree well with the predicted values $-\frac{1}{2}$ and $-\frac{1}{3}$ (see Eqs.~(\ref{eqn_cor_pi}) and (\ref{eqn_cor_C})). The corresponding slopes are shown in dotted lines, as a guide to the eye.}
\label{Cpi_late_time}
\end{figure}

The late time behaviour of the two correlation functions can also be tested numerically. Fig.~\ref{Cpi_late_time} shows $\Pi(t_w+t,t_w)$ versus $t/t_w^{\nu}$, and $C(t_w+t,t_w)$ versus $t/t_w$, for $t_w=10^5$ and $10^3$ respectively, and $\mu=0.5$. The two correlation functions behave as power laws at large time $t$, and the exponents are in good agreement with those predicted by Eq.~(\ref{eqn_cor_pi}) and~(\ref{eqn_cor_C}), shown for comparison (with a arbitrary prefactor), at least for this particular value of $\mu$. Note that since we do not know the constants $\gamma_0$ and $\gamma_{\infty}$, we cannot test the values predicted for the prefactors.

\section{Conclusion}

In this paper, we have studied in details the one dimensional exponential trap model, 
which exhibit a phase transition between a high temperature diffusive phase and a 
low temperature sub-diffusive phase. We have obtained numerically and analytically the
shape of the average diffusion front in the sub-diffusive phase. Although based on an 
approximation valid only near the dynamical transition, our calculation provides several
predictions on the asymptotic shape of $\langle p(x,t) \rangle$ which are in excellent 
agreement with the numerics. It would be interesting to see whether these predictions are
actually exact. 

The central result of this study concerns the localization properties. We have found 
that the dynamical participation ratios are all finite, but different from their 
equilibrium counterparts, even allowing for the existence of an effective, dynamical 
temperature. This is surprising because since each site is visited a
very large number of times by the random walk, one could have expected that a 
partial equilibrium sets in within the limited region of space explored by the walk. 
Our detailed study of the distribution of dynamical weights shows that this is not
the case. We have argued that this can be interpreted in terms of an effective `fragmentation' of space
in two half lines (or even three domains), with a restricted equilibrium within each region, independently of the others.  

Finally, we have studied two different two-time correlation functions, which exhibit 
different aging properties: one -- $\Pi(t_w+t,t_w)$ -- is `sub-aging' whereas the 
other one -- $C(t_w+t,t_w)$ -- shows `full aging'. We have given intuitive arguments and simple 
analytical approximations that account 
for these differences. We have obtained new predictions for the asymptotic (short time and
long time) behaviour of the scaling function associated to $C(t_w+t,t_w)$, which 
are found to be in excellent agreement with the numerics. Since two time scales ($t_w^\nu$ and $t_w$) appear in this model, one can wonder whether the 
short time behaviour of $C(t_w+t,t_w)$ exhibits a non-trivial, multiple time scaling. 
A careful numerical investigation of this issue leads to a negative answer, although 
strong finite time corrections are expected. 

Since this one dimensional model is currently of
interest to the mathematical community, we hope that the study presented 
here will motivate further rigorous research, and that some of our results, in
particular concerning  asymptotic estimates, can be proven to be exact.

\section*{Acknowledgments}
Fruitful discussions on this model with G. Ben Arous, J.-M. Luck, C. Newman, 
P. Maass, M. M\'ezard, C. Monthus and G. de Smedt are gratefully acknowledged.


\section*{Appendix A: The average diffusion front}

\subsection*{Formulation of the problem}

The explicit calculation of $\la p(x,t) \rat$ is reported is this appendix. 
Two different averages will be introduced, the average over the random walks 
$\la \dots \raw$, and the average over the quenched trapping times $\la \dots \rat$. 
We consider here a slightly modified version of the model, in which the particle stays 
on a site a time exactly equal to $\tau(x)$ (continuous notations are used, so as to 
facilitate the continuous space limit) rather than exponentially distributed around 
this value. We expect, and have checked numerically that this is irrelevant for the 
shape of $\la p(x,t) \rat$ at long times.

For a given sample of the disorder (quenched trapping times), we can 
decompose the probability $p(x,t)$ for the walker to be on site $x$ at 
time $t$ into a sum over the number $n$ of steps:

\be
p(x,t) = \sum_{n=0}^{\infty} P(x,n;t) = \sum_{n=0}^{\infty} \la 
\delta_{x_n,x} I(t_n<t<t_{n+1}) \raw
\ee
where $I(t_n<t<t_{n+1})$ is the characteristic function of the interval 
$[t_n,t_{n+1}]$, equal to $1$ of $t$ belongs to this interval, and $0$ otherwise. 
In order to simplify the notations, we introduce $I_n(t) = I(t_n<t<t_{n+1})$. 
Now, averaging over the disorder:
\be
\la p(x,t) \rat = \sum_{n=0}^{\infty} \la \la \delta_{x_n,x} I_n(t) \raw \rat
\ee
The key-point is that we can permute the two averages, and perform first 
the average over the disorder for a given walk. Introducing the average 
$\la \dots \rangle_{n,x}$ over the $n$ steps walks ending on site $x$, we get:
\be
\la p(x,t) \rat = \sum_{n=0}^{\infty} q(x|n) \la \la I_n(t) \rat \rangle_{n,x}
\ee
where $q(x|n)$ is the standard probability for the random walk to 
be on site $x$ after $n$ steps:
\be
q(x|n) = \frac{1}{\sqrt{2 \pi n}}\, e^{-x^2/2n}
\ee
for large $n$. Taking the temporal Laplace transform ${\cal L}$ of $\la p(x,t) \rat$:
\be \label{pxs}
\la \hat{p}(x,s) \rat = \sum_{n=0}^{\infty} q(x|n) \la \la \hat{I}_n(s) \rat \rangle_{n,x}
\ee
which requires the calculation of $\hat{I}_n(s)$:
\bea
\hat{I}_n(s) &=& \int_{t_n}^{t_{n+1}} e^{-st} dt = \frac{1}{s} \, 
e^{-st_n} \, [1-e^{-s(t_{n+1}-t_n)}] \\ \nonumber
\hat{I}_n(s) &\simeq& \tau(x)\, e^{-st_n}
\eea
since $t_{n+1}-t_n = \tau(x)$, and $\tau(x)$ is at most of order $s^{-\nu}$ 
when $s \to 0$, so that $s \tau(x)$ should be small.
For a given walk $W$ ending on site $x$ after $n$ steps, the time $t_n$ can be 
decomposed into a sum over the different visited sites:
\be
t_n = \sum_{x'} {\cal N}_W(x')\, \tau(x')
\ee
where ${\cal N}_W(x')$ is the number of visits of the site $x'$ by the walk $W$. 
Now $\hat{I}_n(s)$ can be averaged over the disorder:
\be
\la \hat{I}_n(s)\rat = \la \tau\, e^{-s {\cal N}_W(x) \tau} \rat \, 
\prod_{x' \neq x} \la e^{-s {\cal N}_W(x') \tau} \rat
\ee
Averages of the form $\la e^{-a\tau} \rat$ or $\la \tau e^{-a\tau} \rat$ are easily calculated, in the limit $a \to 0$:
\bea
\la e^{-a\tau} \rat &=& \int_1^{\infty} \frac{\mu d\tau}{\tau^{1+\mu}} e^{-a\tau} 
\simeq 1 - c\, a^{\mu} \simeq e^{-c a^{\mu}} \label{exp_at}\\
\la \tau e^{-a\tau} \rat &=& - \frac{\partial}{\partial a} \la e^{-a\tau} \rat = c\, \mu\, a^{\mu-1} e^{-c a^{\mu}}
\eea
with $c=\Gamma(1-\mu)$, so that $\la \hat{I}_n(s) \rat$ reads:
\be \label{Ins}
\la \hat{I}_n(s) \rat = c\, \mu \left[ s {\cal N}_W(x) \right]^{\mu-1} 
e^{-c s^{\mu} \left[\sum_{x'}{\cal N}_W(x')^\mu \right]}
\ee
Next, one has to average over all the walks $W$ ending on site $x$ in $n$ steps. 
Since this part is the hardest one of the calculation, one has to resort to a
simple approximation scheme, valid in the vicinity of some specific value 
of $\mu$, namely $\mu$ close to $1$ in the following.

\subsubsection*{An approximation for $\mu \to 1$}

A simple approximation consists in performing the average 
$\la \dots \rangle_{n,x}$ of the right hand side of Eq.~(\ref{Ins}) by replacing ${\cal N}_W(x')$ with ${\cal N}_x(x',n) = 
\la {\cal N}_W(x') \rangle_{n,x}$:
\bea \nonumber
\la \la \hat{I}_n(s) &\rat& \rangle_{n,x} = \la c \mu \left[ s {\cal N}_W(x) \right]^{\mu-1} e^{-c s^{\mu} \left[\sum_{x'}{\cal N}_W(x')^{\mu} \right]} \rangle_{n,x}\\
&\simeq& c \mu \left[ s {\cal N}_x(x,n) \right]^{\mu-1} e^{-c s^{\mu} \left[\sum_{x'}{\cal N}_x(x',n)^{\mu} \right]} \label{pns}
\eea
This approximation is expected to be correct, at large times, for $\mu$ close to $1$ ($\mu<1$), since it is exact for $\mu=1$. Note however that for $\mu=1$, Eq.~(\ref{exp_at}) is no longer valid, and logarithmic corrections come into play. Turning to ${\cal N}_x(x',n)$, it can be shown that for large $n$, a scaling relation holds:
\be
{\cal N}_x(x',n) = \sqrt{n}\, F\left(\frac{x'}{\sqrt{n}}, \frac{x}{\sqrt{n}}\right)
\ee
with $F(v,z)$ given by the following integral:
\be
F(v,z) = \frac{1}{\sqrt{2\pi}} \int_0^1 \frac{du}{\sqrt{\sin \pi u}}\, e^{-\frac{(v-zu)^2}{2 \sin \pi u}}
\ee
The different factors in Eq.~(\ref{pns}) can then be evaluated:
\bea
\sum_{x'}{\cal N}_x(x',n)^{\mu} &=& \int_{-\infty}^{+\infty} \sqrt{n} dv \left[ \sqrt{n} F\left(v,\frac{x}{\sqrt{n}}\right) \right]^{\mu} \\ \nonumber
&=& \sqrt{n}^{1+\mu} G\left(\frac{x}{\sqrt{n}}\right)
\eea
introducing $G(z)=\int_{-\infty}^{+\infty} F(v,z)^{\mu} dv$.
\bea
{\cal N}_x(x,n)^{\mu-1} &=& \sqrt{n}^{\mu-1} F\left(\frac{x}{\sqrt{n}},\frac{x}{\sqrt{n}}\right)^{\mu-1} \\ \nonumber
&=& \sqrt{n}^{\mu-1} H\left(\frac{x}{\sqrt{n}}\right)^{\mu-1}
\eea
with $H(v)=F(v,v)$. From Eq.~(\ref{pxs}), a continuous space limit can be obtained, introducing a continuous scaling variable $\lambda$ through the natural scaling relation $n=\lambda x^2$:
\bea
\la \hat{p}(x,s) \rat &=& x^2 \int_0^{\infty} d\lambda \left[ \frac{e^{-1/2\lambda}}{|x| \sqrt{2\pi \lambda}}\right]\\ \nonumber
&\times& \, \left[c \mu s^{\mu-1} (|x| \sqrt{\lambda})^{\mu-1} H\left(\frac{1}{\sqrt{\lambda}}\right)^{\mu-1} \right] \\ \nonumber
&\times& \, e^{-\left[c s^{\mu} |x|^{1+\mu} \sqrt{\lambda}^{1+\mu} G\left(\frac{1}{\sqrt{\lambda}}\right)\right]}
\eea
where the parity of $G(z)$ and $H(z)$ has been used. Grouping together the factors, and introducing the scaling variable $\eta = |x|^{(1+\mu)/\mu} s$, we get:
\bea \label{pxscont}
\la \hat{p}(&x&,s) \rat = |x|^{1/\mu}\, \frac{\mu c}{\sqrt{2\pi}}\, \eta^{\mu-1} \int_0^{\infty} d\lambda\, \lambda^{\frac{\mu}{2}-1} \\ \nonumber
&\times& H\left(\frac{1}{\sqrt{\lambda}}\right)^{\mu-1}\, e^{-\frac{1}{2\lambda} - c \eta^{\mu} \lambda^{(1+\mu)/2} G\left(\frac{1}{\sqrt{\lambda}}\right)}
\eea
Note that this expression is compatible with the expected scaling form: 
\be \label{gscal}
\la p(x,t) \rat = \frac{1}{\xi}\, f\left(\frac{|x|}{\xi}\right)
\ee
where $\xi$ is the dynamical length scale appearing in the model: $\xi \sim t^{\mu/(1+\mu)}$. One can indeed rewrite the previous relation in the following way:
\be
\la p(x,t) \rat = \frac{1}{|x|}\, g\left(\frac{t}{|x|^{(1+\mu)/\mu}}\right)
\ee
Taking the Laplace transform with respect to $t$ yields:
\be
\la \hat{p}(x,s) \rat = |x|^{1/\mu} \hat{g}(\eta)
\ee
where $\hat{g}$ is the Laplace transform of $g$, which is of the form
 Eq.~(\ref{pxscont}). So one deduces that the scaling function $\hat{g}(\eta)$ is given by:
\bea \label{feta0}
&\hat{g}&(\eta) = \frac{\mu c}{\sqrt{2\pi}}\, \eta^{\mu-1} \int_0^{\infty} d\lambda\, \lambda^{\frac{\mu}{2}-1} \\ \nonumber
&\times& H\left(\frac{1}{\sqrt{\lambda}}\right)^{\mu-1}\, e^{-\frac{1}{2\lambda} - c \eta^{\mu} \lambda^{(1+\mu)/2} G\left(\frac{1}{\sqrt{\lambda}}\right)}
\eea

\subsubsection*{Asymptotic behaviour of the scaling function $f$}

Now we focus on the asymptotic behaviour of $\hat{g}(\eta)$ for large $\eta$, which gives the spatial tails of the distribution $\la p(x,t) \rat$. When $\eta \to \infty$, the above integral is dominated by the small $\lambda$ region, which means that one needs to know the asymptotic large $z$ behaviour of $H(z)$ and $G(z)$. After a
few lines of computations, we finally find:
\be
H(z) \simeq \frac{1}{z} \qquad
G(z) \simeq z^{1-\mu}    \qquad z \to \infty
\ee
The large $\eta$ behaviour of $\hat{g}(\eta)$ can be then obtained from:
\be
\hat{g}(\eta) = \frac{\mu c}{\sqrt{2\pi}}\, \eta^{\mu-1} \int_0^{+\infty} d\lambda\, \lambda^{\mu-\frac{3}{2}} e^{-\frac{1}{2\lambda}-c \eta^{\mu} \lambda^{\mu}}.
\ee
The inverse Laplace transform can be computed using a saddle-point method.
One finally finds for the large $|x|$ behaviour (or large $|\zeta|$, with $\zeta=xt^{-\frac{\mu}{1+\mu}}$):
\be
f(\zeta) \approx  f_\infty
|\zeta|^{\frac{\mu-1}{2}} e^{- b |\zeta|^{1+\mu}}
\ee
with:
\be\label{factor1}
f_\infty= \sqrt{\frac{\mu \Gamma(1-\mu)}{2^{\mu}\pi}}, \qquad b=2^{-\mu} \Gamma(1-\mu).
\ee

One can also look at the limit $\zeta \to 0$. Starting from Eq.~(\ref{feta0}), we have to calculate the small $z$ behaviour of $G(z)$ and $H(z)$, which is simple here since these function have a finite limit in $0$, denoted by $g_0(\mu)$ and $h_0$ respectively:
\bea
g_0(\mu) &=& \int_{-\infty}^{+\infty} dv \left( \frac{1}{\sqrt{2\pi}} \int_0^1 \frac{du}{\sqrt{\sin \pi u}}\, e^{\frac{v^2}{2\sin \pi u}} \right)^{\mu}\\
h_0 &=& \frac{1}{\sqrt{2\pi}} \int_0^1 \frac{du}{\sqrt{\sin \pi u}} = \frac{\Gamma \left(\frac{1}{4}\right)^2}{2 \pi^2}
\eea
So we have to compute the following integral:
\be
\hat{g}(\eta) = \frac{\mu c}{\sqrt{2\pi}}\, h_0^{\mu-1} \eta^{\mu-1} \int_0^{\infty} d\lambda\, \lambda^{\frac{\mu}{2}-1} e^{-\frac{1}{2\lambda}-c g_0(\mu) \eta^{\mu} \lambda^{(1+\mu)/2}}
\ee
For $\eta \to 0$, this integral is dominated by the large $\lambda$ region. We 
finally find:
\be
\hat{g}(\eta) = A\, \eta^{-\frac{1}{1+\mu}} \qquad \eta \to 0
\ee
with:
\be
A=\frac{1}{\sqrt{2\pi}} \frac{2\mu}{1+\mu} \Gamma(1-\mu)^{\frac{1}{1+\mu}} g_0(\mu)^{-\frac{\mu}{1+\mu}} h_0^{\mu-1} \Gamma\left(\frac{\mu}{1+\mu}\right)
\ee
Taking the inverse Laplace transform, this term gives the value of
$\la p(x=0,t)\rat$ which is proportional to $t^{-\frac{\mu}{1+\mu}}$. In order to get the spatial dependence, the next term of the expansion must be computed. 
After a few changes of variables and asymptotic estimates, we get:
\be
f(\zeta) = f_0 - f_1 |\zeta|^{\mu} \qquad \zeta \to 0
\ee
with:
\bea \label{factor2}
f_0 &=& \frac{2\mu h_0^{\mu-1}}{(1+\mu)\sqrt{2\pi}} g_0(\mu)^{-\frac{\mu}{1+\mu}} \frac{\Gamma(1-\mu)^{\frac{1}{1+\mu}} \Gamma \left(\frac{\mu}{1+\mu}\right)}{\Gamma \left(\frac{1}{1+\mu} \right)} \\
f_1 &=& \frac{2^{\frac{1-\mu}{2}}}{\sqrt{\pi}} h_0^{\mu-1} \Gamma \left( 1-\frac{\mu}{2}\right)
\eea
The next sub-leading term can also be computed, and is found to be of order $\zeta^2$
for $\mu > 1/2$, and $\zeta^{1+2\mu}$ for $\mu < 1/2$.

\section*{Appendix B}

We give here some technical details about the numerical simulations. A number $N_w$ of independent `walkers' (or `particles') are simulated one by one, for a given sample of the quenched energies $\{E_i\}$. A walk is simulated as follows: the trapping time on site $i$ is chosen randomly from an exponential distribution of mean $\tau_i = \exp(E_i/T)$, and the walker then chooses at random between the two neighbouring sites, with equal probability. Then the desired quantity is computed for this particular sample, and eventually averaged over a number $N_s$ of samples. Moreover, in order to facilitate comparisons between different runs, we took each time the same disorder samples, by choosing the same set of `seed' numbers. For out of equilibrium simulations, where the number of sites $L=2N+1$ is supposed to be infinite, we used periodic boundary conditions, with usually $N=10^3$ except when long times were required, in which case $N=10^4$ was instead considered (for instance in the computation of $Y_k(t)$).

Error bars are estimated by running several simulations with the same $N_w$ and $N_s$, varying only the `seed' numbers. The fluctuations between the different runs leads to an estimate of the standard deviation. The numbers $N_w$ and $N_s$ were chosen so as to get small enough error bars, as far as possible, taking into account the time $t$ that we need to reach as well as the computational time. It should be emphasized that the amplitude of the fluctuations depends a lot on the computed quantity. Correlations functions are easy to compute, and $N_w=N_s=10^3$ is enough to get a relative standard deviation which is less than $10^{-2}$. Quantities related to localization properties fluctuate more; for integrated quantities like the participation ratios, we took $N_w=10^3$ and $N_s=10^4$ (except for small size systems where $N_w=N_s=2.10^3$ was used instead), which was a good compromise in order to obtain both a good standard deviation (less than $10^{-2}$) and long enough times (for instance $t=10^6$ for $\mu=0.5$). For distributions like $\varphi(P)$ and $\la p(x,t) \ra$, one needs to average over a larger number of samples, in order to get smooth enough curves (with fluctuations between different runs less than $5.10^{-2}$). So  $\varphi(P)$ was simulated using $N_w=10^4$ and $N_s=10^5$, whereas $\la p(x,t) \ra$ was computed with $N_w=10^3$ and $N_s=10^5$. Note finally that $Y_2(\ell,t)$ was simulated using $N_w=10^5$, and $N_s=10^3$ (except for $t=10^6$ and $10^7$, where $N_w=5.10^4$ and $10^4$ respectively), since we need a good statistics on the sites with small trapping times so as to avoid large fluctuations at small $\ell$.

\section*{Appendix C: Calculation of the participation ratios}

The analytical calculation of $Y_2(t)$ appears not to be easily tractable, and the aim of this appendix is to argue for the existence of a finite limit $Y_2^{dyn}$ of $Y_2(t)$ when $t \to \infty$ for $\mu < 1$, and try to extract some information on the behaviour of $Y_2^{dyn}$ as a function of $\mu$, in particular for $\mu$ close to $1$.

The participation ratio $Y_2(t)$ is given by the integral 
$\int_{-\infty}^{+\infty} \la p(x,t)^2 \rat$. The quantity $\la p(x,t)^2 \rat$ 
can be computed following the same lines as for $\la p(x,t) \rat$. It will be useful to introduce a two-time quantity $Q(x,t,t')$ defined as:
\be
Q(x,t,t') = \la p(x,t) p(x,t') \rat
\ee
Defining $R(t,t')=\int_{-\infty}^{\infty} dx\, Q(x,t,t')$, one has $Y_2(t)=R(t,t)$, and turning to the Laplace transform:
\be
\hat{R}(s,s')= \int_0^{\infty} dt \int_0^{\infty} dt' e^{-st-s't'} R(t,t')
\ee
Now a reasonable assumption (that has been checked numerically) is that for large $t$ and $t'$, $t>t'$, $R(t,t')$ becomes a function of $\frac{t}{t'}$:
\be
R(t,t') = Y_2^{dyn} {\cal R}\left(\frac{t}{t'}\right)
\ee
This follows from the similar behaviour of the correlation function $C(t,t')$ studied in section \ref{sect-correl}.
It is interesting to restrict to the particular case $s=s'$:
\bea
\hat{R}(s,s) &=& 2\, Y_2^{dyn} \int_0^{\infty} dt' \int_t'^{\infty} dt\, e^{-s(t+t')} {\cal R}\left(\frac{t}{t'}\right) \\ \nonumber
&=&  2\, Y_2^{dyn} \int_0^{\infty} dt' \int_1^{\infty} du \,t' e^{-st'(1+u)} {\cal R}(u)
\eea
or:
\be
\hat{R}(s,s)= \frac{2 Y_2^{dyn}}{s^2} \int_1^{\infty} du \frac{{\cal R}(u)}{(1+u)^2}
\ee
So a finite limit for $Y_2(t)$ corresponds to $\hat{R}(s,s) \sim s^{-2}$, and this is what we shall try to show in the following.
Coming back to $Q(x,t,t')$ and decomposing over the number of steps, one has:
\be
Q(x,t,t') = \sum_{n,n'} p(x,n,t)\, p(x,n',t')
\ee
Averaging over the disorder yields:
\bea
\la &Q&(x,t,t') \rat = \sum_{n,n'} q(x|n) q(x|n') \\ \nonumber
&\times& \la I(t_n<t<t_{n+1}) I(t'_{n'}<t'<t'_{n'+1}) \ra_{\tau, (n,x), (n',x)}
\eea
For given walks $W$ and $W'$, and a given sequence of $\tau_i$, let us introduce $\hat{K}_{n,n'}(s,s')$ defined by:
\bea
&\hat{K}_{n,n'}&(s,s') = \int_0^{\infty} dt \int_0^{\infty} dt' I(t_n<t<t_{n+1}) \\ \nonumber
&\times& I(t'_{n'}<t'<t'_{n'+1})\, e^{-st-s't'} 
\simeq \tau(x)^2 e^{-st_n-s't'_{n'}}
\eea
assuming again that $t_{n+1}-t_{n}=t'_{n'+1}-t'_{n'}=\tau(x)$ (i.e. trapping times are fixed rather than exponentially distributed) and that $s\tau(x)$ and $s'\tau(x)$ are both much smaller than $1$, which means that the maximum trapping time encountered is much smaller than the times $t$ and $t'$ considered. Introducing the following decomposition:
\bea
t_n &=& \sum_y {\cal N}_W(y,n)\, \tau(y)\\
t'_{n'} &=& \sum_y {\cal N}_{W'}(y,n')\, \tau(y)
\eea
$\hat{K}_{n,n'}(s,s')$ reads:
\bea
\hat{K}_{n,n'}(s,s') &=& \tau(x)^2 e^{-[s{\cal N}_W(x,n)+s'{\cal N}_{W'}(x,n')]\tau(x)} \\ \nonumber
&\times& \prod_{y\neq x} e^{-[s{\cal N}_W(y,n)+s'{\cal N}_{W'}(y,n')]\tau(y)}
\eea
Averaging $\hat{K}_{n,n'}(s,s')$ over the disorder, one has:
\bea
\la \hat{K}_{n,n'}(s,s') \rat &=& \mu \Gamma(2-\mu) \\ \nonumber
&\times& \left[s{\cal N}_W(x,n)+s'{\cal N}_{W'}(x,n')\right]^{\mu-2} \\ \nonumber &\times& \prod_{y \neq x} e^{-c \left[s{\cal N}_W(y,n)+s'{\cal N}_W'(y,n')\right]^{\mu}}
\eea
One can now write $\hat{R}(s,s)$ as:
\bea
\hat{R}(s,s) &=& 2\mu \Gamma(2-\mu) \sum_x \sum_{n<n'} q(x|n) q(x|n') s^{\mu-2} \\ \nonumber
&\times& \left[{\cal N}_W(x,n)+{\cal N}_{W'}(x,n') \right]^{\mu-2} \\ \nonumber
&\times& e^{-c s^{\mu} \sum_y \left[{\cal N}_W(y,n)+{\cal N}_{W'}(y,n') \right]^{\mu}}
\eea
We now turn to continuous limit, and replace as in Appendix A ${\cal N}_W(y,n)$ by its average value $\sqrt{n} F\left(\frac{y}{\sqrt{n}},\frac{x}{\sqrt{n}}\right)$. At this stage we drop order unity constants since we shall make several rather crude approximations in the following. Introducing the new variable $\beta$ through $n'=\beta n$, one gets:
\bea
\hat{R}(s,s) &\sim& s^{\mu-2} \int_{-\infty}^{\infty} dx \int_1^{\infty} dn \int_1^{\infty} \frac{d\beta}{\sqrt{\beta}} e^{-\frac{x^2}{2n}(1+\frac{1}{\beta})} \\ \nonumber
&\times& \left[\sqrt{n} H\left(\frac{x}{\sqrt{n}}\right) + \sqrt{\beta n} H\left(\frac{x}{\sqrt{\beta n}}\right) \right]^{\mu-2} \\ \nonumber
&\times& e^{-c s^\mu \int_{-\infty}^{\infty} dy\, n^{\frac{\mu}{2}} \left[F\left(\frac{y}{\sqrt{n}},\frac{x}{\sqrt{n}}\right)+\sqrt{\beta} F\left(\frac{y}{\sqrt{\beta n}},\frac{x}{\sqrt{\beta n}}\right)\right]^{\mu}}
\eea
Now, because of the factor $\frac{1}{\sqrt{\beta}}$, the integral over $\beta$ is dominated by the large $\beta$ behaviour, which means that as a first step one can neglect terms like $F\left(\frac{y}{\sqrt{n}},\frac{x}{\sqrt{n}}\right)$ compared to $\sqrt{\beta} F\left(\frac{y}{\sqrt{\beta n}},\frac{x}{\sqrt{\beta n}}\right)$. Rescaling also $x$ and $y$ using the new variables $\hat{x}=\frac{x}{\sqrt{n}}$ and $\hat{y}=\frac{y}{\sqrt{\beta n}}$ yields:
\bea
\hat{R}(s,s) &\sim& s^{\mu-2} \int_{-\infty}^{\infty} d\hat{x} \int_1^{\infty} d\beta\, \beta^{\frac{\mu-3}{2}} \int_1^{\infty} dn\, n^{\frac{\mu-1}{2}} \\ \nonumber
&\times& e^{-\frac{\hat{x}}{2}(1+\frac{1}{\beta})}
 H\left(\frac{\hat{x}}{\sqrt{\beta}}\right)^{\mu-2} e^{-cs^{\mu}(\beta n)^{\frac{1+\mu}{2}} G\left(\frac{\hat{x}}{\sqrt{\beta}}\right)}
\eea
One can change variable in the last integral over $n$, letting:
\be
n = \frac{1}{\beta s^{\frac{2\mu}{1+\mu}}} \left[\frac{v}{c \,G(\frac{\hat{x}}{\sqrt{\beta}})}\right]^{\frac{2}{1+\mu}}\\
\ee
Then the integral over $v$ becomes $\int_0^{\infty} dv\, e^{-v} = 1$ (for $s \to 0$), and $\hat{R}(s,s)$ reduces to:
\be
\hat{R}(s,s) \sim \frac{1}{c\, s^2} \int_1^{\infty} \frac{d\beta}{\beta^2} \int_{-\infty}^{\infty} d\hat{x}\, e^{-\frac{\hat{x}}{2}(1+\frac{1}{\beta})} \frac{H\left(\frac{\hat{x}}{\sqrt{\beta}}\right)^{\mu-2}}{G\left(\frac{\hat{x}}{\sqrt{\beta}}\right)}
\ee
So this simplified calculation is consistent with a finite $Y_2^{dyn}$ given by:
\be
Y_2^{dyn} \propto \frac{1}{\Gamma(1-\mu)} \int_1^{\infty} \frac{d\beta}{\beta^2} \int_{-\infty}^{\infty} d\hat{x}\, e^{-\frac{\hat{x}}{2}(1+\frac{1}{\beta})} \frac{H\left(\frac{\hat{x}}{\sqrt{\beta}}\right)^{\mu-2}}{G\left(\frac{\hat{x}}
{\sqrt{\beta}}\right)}
\ee
where we have used $c=\Gamma(1-\mu)$.

Since $\lim_{z \to 0} G(z) = g_0(\mu) \to 1$ when $\mu \to 1$ and 
$\lim_{z \to 0} H(z) = h_0$ is independent of $\mu$, the only strong dependence upon $\mu$ comes from $\Gamma(1-\mu)$, which suggests that:
\be
Y_2^{dyn} \sim (1-\mu) \qquad \mu \to 0
\ee
This results is compatible with numerical data, and is comparable to the corresponding 
result in equilibrium (in this case, $Y_2^{eq}=1-\mu$ for all $\mu<1$).
For $1<\mu<2$, one can easily show that, under the same assumptions, $Y_2(t) \sim 
1/t^{(\mu-1)/2}$ when $t \to \infty$, whereas $Y_2(t) \sim 
1/\sqrt{t}$ for $\mu > 2$, which is the expected result. 
Turning to $Y_3$, we have checked that the same calculation also leads to a finite limit, and to a linear behaviour with respect to $\mu$ when $\mu \to 1$.

\end{multicols}

\end{document}